\documentclass[amsmath,nofootinbib]{revtex4-1}
\usepackage{graphics,setspace,epsfig,color}
\usepackage{epsf}
\usepackage{amscd}
\usepackage{amsmath}
\usepackage{amsfonts}
\usepackage{microtype}

\newcommand{\Real}{\mathbb{R}}

\newcommand{\beq}{\begin{equation}}
\newcommand{\eeq}{\end{equation}}
\newcommand{\bea}{\begin{eqnarray}}
\newcommand{\eea}{\end{eqnarray}}
\newcommand{\nn}{\nonumber}
\newcommand{\eq}[1]{Eq.~\ref{#1}}
\newcommand{\fig}[1]{Fig.~\ref{#1}}
\newcommand{\tab}[1]{Table~\ref{#1}}

\def\beqs#1\eeqs{\beq\begin{split} #1 \end{split}\eeq}
\def\dd#1#2{\frac{d #1}{d #2}}
\def\pd#1#2{\frac{\partial #1}{\partial #2}}
\let\Re\relax 
\DeclareMathOperator\Re{Re}
\let\Im\relax 
\DeclareMathOperator\Im{Im}

\usepackage{multirow}
\usepackage{booktabs}

\begin{document}

\title{A Monte Carlo algorithm for simulating fermions on Lefschetz thimbles}
\author{ Andrei Alexandru}
\affiliation{Department of Physics \\
The George Washington University \\
Washington, DC 20052}
\author{G\"ok\c ce Ba\c sar}
\author{ Paulo Bedaque }
\affiliation{Department of Physics \\
University of Maryland\\College Park, MD 20742}


\begin{abstract}
A possible solution of the notorious sign problem preventing direct
Monte Carlo calculations for systems with non-zero chemical potential is to deform
the integration region in the complex plane to a Lefschetz thimble. We investigate
this approach for a simple fermionic model. We introduce an easy to implement 
Monte Carlo algorithm to sample the dominant thimble. Our algorithm relies 
only on the integration of the gradient flow in the numerically stable direction, 
which gives it a distinct advantage over the other proposed algorithms. We demonstrate 
the stability and efficiency of the algorithm by applying it to an exactly solvable fermionic 
model and compare our results with the analytical ones. We report a
very good agreement for a certain region in the parameter space where the 
dominant contribution comes from a single thimble, including a region where standard
methods suffer from a severe sign problem. However, we find that there are also 
regions in the parameter space where the contribution from multiple thimbles is important,
even in the continuum limit.

\end{abstract}

\maketitle

\section{Introduction}

The Monte Carlo study of many field theories and many-body systems is impeded by the sign problem. Among those theories are  theories of great importance such as QCD at finite density (central to nuclear physics and the physics of neutron stars) and the Hubbard model (important in the theory of high temperature superconductors).  In the language of field theory, where observables of the theory are expressed as a path integral, the sign problem appears as the fact that the integrand is highly oscillatory and delicate cancellations between contributions with opposite signs are required to produce the correct result. Due to the importance of this problem several methods have been proposed and explored aiming at its solution. Those include series expansions on the chemical potential~\cite{Allton:2002zi}, re-weighting~\cite{Fodor:2001au,Fodor:2004nz}, canonical partition function methods~\cite{Alexandru:2005ix,deForcrand:2006ec}, analytical continuation from imaginary chemical potential~\cite{deForcrand:2006pv}, and complex Langevin/stochastic quantization~\cite{Aarts:2008rr}. 

Recently there has been progress in implementing a new method~\cite{Cristoforetti:2012su,Cristoforetti:2013wha,Fujii:2013sra,Mukherjee:2013aga,Cristoforetti:2013qaa,Cristoforetti:2014gsa,Aarts:2014nxa,Tanizaki:2014tua,Tanizaki:2015rda,Fukushima:2015qza,Tsutsui:2015tua,Fodor:2015doa,Fujii:2015bua,Fujii:2015vha}. The basic idea is to complexify the real variables of the path integral and deform the region of integration to a region where the integrand is real and positive. This new region of integration is a generalization of the steepest descent path to multidimensional spaces. It can be expressed as a combination of certain integration regions, each of which is attached to a critical point of the action (i.e. the configurations $A$ such that $\delta S[A]/\delta A=0$). These integration regions are known as "Lefschetz thimbles" and the integrand is real and positive (up to an overall phase) over each thimble. The original integral is then expressed as a combination of integrals over those thimbles. This idea has ben fruitful in different aspects of quantum mechanics~\cite{Witten:2010zr,Behtash:2015kna,Behtash:2015kva,Behtash:2015zha} and quantum field theories \cite{Witten:2010cx,Harlow:2011ny,Cherman:2014ofa}, especially in studying semiclassical expansions. In this paper, however, we will focus on the implementation of the Lefschetz thimble approach on lattice field theory.

In general, expressing multi-dimensional, complex, Laplace type integrals in terms these Lefschetz thimbles is governed by a well-developed mathematical theory (Picard-Lefschetz theory)~\cite{fedoryuk,pham,kaminski}. This theory is, however, difficult to apply in practice. Especially for integrals that appear in lattice field theory, finding out which particular set of thimbles contributes to the original path integral is a difficult problem.  Some arguments have been made~\cite{Cristoforetti:2012su,Cristoforetti:2013wha} that the theory defined over one single thimble (associated with the perturbative vacuum) is in the same universality class as the original one. This is not a rigorous argument and a lot of doubt remains, specially in the case of fermionic theories. Furthermore, it is unclear whether Picard-Lefschetz theory applies to the case where the action is not  a polynomial, as in the case of fermionic theories.\footnote{ Notably there are examples with non-polynomial actions, even actions with poles where the theory applies \cite{Basar:2013eka,Cherman:2014ofa}.  }
This motivated the analytical study of toy models for the fermionic case~\cite{Fujii:2015bua,Kanazawa:2014qma,Tanizaki:2015rda} that shed some of light of these issues.

Different algorithms have been proposed to compute integrals over a thimble and then applied to bosonic theories\footnote{At the final stages of this research the first application to a fermionic model appeared \cite{Fujii:2015bua,Fujii:2015vha}.}.
In the present paper we propose a novel algorithm to compute integrals over one thimble and test it on a simple, exactly solvable,  $0+1$ dimensional fermionic theory. Unlike some of the previously introduced algorithms, our algorithm does not suffer from certain problems such as the unstable flow towards the critical point. We verify the feasibility and efficiency of the algorithm, by comparing our results with the exact ones, and find regions of the parameters space where the contribution of other thimbles is negligible. Those regions include models with a severe sign problem that could not be dealt with using mode standard Monte Carlo methods and successfully reproduces the exact result with a good precision. On the other hand, we also argue that there are other regions in the parameter space where the integration over one thimble is not sufficient and contributions from other thimbles have to be included. The existence of these regions persists in the continuum limit. The fermionic determinant which leads to questions on the applicability of the Picard-Lefschetz theory to systems with fermions does not seem to create any problems for our algorithm.

The organization of the paper is as follows: In Section \ref{sec:thimbles} we recapitulate the basics of Lefschetz thimbles focusing on their implementation on lattice field theory. In Section \ref{sec:algorithm} we explain our Monte Carlo algorithm for computing integrals over one thimble. Section \ref{sec:model} contains a summary of the fermionic model with nonzero chemical potential that we use and a collection of the exact results that we compare our Monte Carlo results to. Our results are given in Section \ref{sec:results}.  We give our conclusions in the final section \ref{sec:conclusions}.

\section{Lefschetz thimbles in lattice field theory}
\label{sec:thimbles}
Consider a multidimensional integral over the real variables $x_i$ (with $i=1,\ldots, N$) as found in the computation of expectation values in euclidean lattice field theory:
\beq
\langle \mathcal{O}[x] \rangle = \frac{\int \! dx\ e^{-S[x]}  \mathcal{O}[x]}{\int \! dx\, e^{-S[x]} }\quad\text{with}\quad dx=\prod_i dx_i.
\eeq When the euclidean action is real this ratio can be estimated  by sampling the $x$-space according to the probability density 
\beq
P[x]=\frac1Ze^{-S[x]}\quad\text{with}\quad Z={\int \! dx\ e^{-S[x]} } \,,
\eeq and computing
\beq
\langle \mathcal{O}[x] \rangle   
\approx
  \frac{1}{N_\text{conf}} \sum_{a=1}^{N_\text{conf}} \mathcal{O}[x^{(a)}] .
\eeq This  method cannot be used if the action $S[x]$ is not real as $P[x]$ is  not a real, positive quantity and cannot be seen as a probability density. This is the generic case for systems, including QCD, with nonzero chemical potential.  An alternative is to split $S$ into its real and imaginary part, include the real part in the probability density and the imaginary part in the observable:
\beqs
\label{eq:reweighting}
\langle \mathcal{O}[x] \rangle &= 
\frac1Z{\int \! dx\ e^{-S[x]}  \mathcal{O}[x]} =
\frac{\int \! dx\ e^{-S_R[x]}  e^{-i S_I[x]} \mathcal{O}[x]}
{\int \! dx\ e^{-S_R[x]}  e^{-i S_I[x]}} \\
&=
\frac{\int \! dx\ e^{-S_R[x]}  e^{-i S_I[x]} \mathcal{O}[x]}
{\int \! dx\ e^{-S_R[x]}} \Bigg/      
\frac{\int \! dx\ e^{-S_R[x]}  e^{-i S_I[x]}}    
{   \int \! dx\ e^{-S_R[x]}           }
=
\frac{  \langle  e^{-i S_I}  \mathcal{O}    \rangle_R}{\langle  e^{-i S_I}\rangle_R}
\,.
\eeqs
The averages $\langle\cdot\rangle_R$ are performed with respect to the positive definite
measure given by the Boltzman factor $\exp(-S_R[x])$. 
The ``reweighting" method described above is successful as long as the fluctuations on the phase $\exp(-i S_I[x])$ are small. It turns out that the phase fluctuation increases exponentially with the spacetime volume and it is of limited practical use.

An alternative is to modify the domain of integration. In one dimension the integration over the real line can be deformed into an integral over a different contour, as long as no singularity is crossed in the deformation process. This feature is explored, for instance, in the steepest descent method where one deforms the contour of integration from the real line to a curve $z(\tau)$ passing through a critical point (a point $z_\text{cr}$ in the complex plane where $dS/dz=0$).
This curve has two properties: i) it is the path along which the real part of the action increases the fastest and ii) the imaginary part of the action is constant along it.
Property i)  makes the choice of the steepest descent path  convenient when performing a semiclassical expansion around the critical point (but, of course, the full exact integral over the new contour equals exactly the original integral). For bypassing the sign problem, however,  property ii) is the useful one. Indeed, since the action has a fixed phase along the new contour, standard Monte Carlo methods can be applied to the evaluation of the integral along the steepest descent path. This curve is the one dimensional version of a Lefschetz thimble. In general, the action can have multiple critical points and therefore multiple thimbles, each of which is attached to a different critical point. The original integral over the real axis is then equal to a particular sum over integrals over certain thimbles. Which thimbles contribute to the original integral depends on the action. How to apply Monte Carlo methods to multi-thimble integrals is an important open question.  

We denote with ${\cal J}$ the integration
contour corresponding to the steepest descent method and $z_\text{cr}$ the critical point
that corresponds to this curve.
In one dimension the curve ${\cal J}$ defining the new integration contour is completely
determined by the two properties listed above. To generalize this construction to 
higher dimension we need to define this curve using the flow induced by the action.
The {\em downward flow}\footnote{Note we use the nomenclature ``downward" in accordance with the direction in which $S_R$ is decreasing.}is a map from of the complex plane into itself, 
$F_\tau:\mathbb{C}\to\mathbb{C}$, with $z_0\mapsto F_\tau(z_0)=z(\tau)$, 
where $z(\tau)$ is the solution of the differential equation:
\beq\label{eq:flow}
\frac{dz}{d\tau} = - \left(  \frac{dS}{dz}  \right)^* \quad\text{with initial condition}\quad
z(0) = z_0 \,.
\eeq 
Using this flow we define the new integration contour ${\cal J}$ passing through the
critical point $z_\text{cr}$
\beq
{\cal J} = \left\{ z\in\mathbb{C}\, \middle|\, \lim_{\tau\to\infty} F_\tau(z) = z_\text{cr} \right\} \,,
\eeq
that is the collection of all points in the plane that flow into the critical point
$z_\text{cr}$.

Before discussing the generalization of these ideas to multi-dimensional integrals, 
we want to stress a few important points. 
\begin{itemize}
\item[-] Starting from any point in the set $z_0\in {\cal J}$, $z_0\not=z_\text{cr}$, 
the flow given by Eq.~\ref{eq:flow} will describe a trajectory in ${\cal J}$ approaching
$z_\text{cr}$ asymptotically. The flow will not cross to the other side of
the critical point.

\item[-] Starting with any point in 
$z_0\in{\cal J}$, the set of 
points ${\cal R}_{z_0}=\{ F_\tau(z_0) | \tau\in(-\infty,\infty)\}$ define a subset of the integration contour (call it the {\em ray} passing through $z_0$). To generate the entire set we need to pick another point in 
$z_1\in{\cal J}$ that is on the other side of the critical point. We have then 
${\cal J}={\cal R}_{z_0} \cup \{z_\text{cr}\} \cup {\cal R}_{z_1}$, where ${\cal R}_{z_1}$
is defined in the same manner as ${\cal R}_{z_0}$. The set ${\cal J}$ is then the union of
all rays, since $z(\tau)=z_\text{cr}$ is also a solution of the flow equation.

\item[-] In one dimension, in order to construct the set ${\cal J}$ as a union of rays, we need two seed points in 
${\cal J}$. These points can be determined by analytically solving the problem in an 
infinitesimal neighborhood of $z_\text{cr}$, where the action is well approximated by
$S[z]\approx S[z_\text{cr}]+\partial^2 S/\partial z^2[z_\text{cr}](z-z_\text{cr})^2/2$.
Around $z_\text{cr}$ the set ${\cal J}$ is approximated by a straight line with the
slope controlled by the phase $\phi = \arg\partial^2 S/\partial z^2[z_\text{cr}]$. As seed
points we can pick two points on this line on each side of $z_\text{cr}$, 
$z_{0,1}=z_\text{cr}\pm\epsilon \exp(-i\phi/2)$, and compute the rays associated with them. The set 
${\cal J}$ is recovered in the limit $\epsilon\to 0$, but in practice a small $\epsilon$ value
is sufficient.

\item[-] Note that if we start with the seed points mentioned above,
 the set of points $z(\tau)$
for $\tau>0$ are all in the neighborhood of $z_\text{cr}$, on the nearly straight line connecting $z_{0,1}$ to $z_\text{cr}$. The points of interest are those for $\tau\in(-\infty,0]$.
In practice this part of the ray is determined using the {\em upward flow}, $\overline{F}_\tau$,
$z_0\mapsto \overline{F}_\tau(z_0)$ defined by the differential equation 
\beq
\label{eq:upflow}
\dd z\tau = \left( \dd S z \right)^* \quad\text{with}\quad z(0)=z_{0} \,.
\eeq
For $\tau>0$ this flow generates the same points as the downward flow $F_\tau$ for 
negative values of $\tau$. In fact the flows are invertible and we have 
$F_\tau^{-1}=F_{-\tau}=\overline{F}_\tau$.

\item[-] We note that there is another curve ${\cal K}$ that passes through the critical
point, along which the imaginary part of the action is constant and equal to 
$S_I[z_\text{cr}]$:
\beq
{\cal K} = \left\{ z\in\mathbb{C}\, \middle|\, \lim_{\tau\to\infty} \overline{F}_\tau(z) = z_\text{cr} \right\} \,.
\eeq
This set is called the {\em unstable thimble} and around the critical point it represents
the direction in which the real part of the action decreases the fastest. The existence
of this set has to do with the fact that the critical points for $S$ are saddle points for
$S_R$.

\item[-] Both sets ${\cal J}$ and ${\cal K}$ are invariant under both the upward and
downward flow, that is points in ${\cal J}$ are moved by both of these flows into points
in ${\cal J}$ (and similarly for ${\cal K}$). From a numerical stability point of view,
we note that the positive upward flow $\overline{F}_{\tau>0}$ is stable on ${\cal J}$, 
that is points that are slightly off ${\cal J}$ are moved by the upward flow into points 
in the vicinity of ${\cal J}$. The positive downward flow $F_{\tau>0}$ is unstable 
around ${\cal J}$ and it is very difficult to integrate the flow in this 
direction numerically.

\end{itemize}

The generalization of these ideas to the multidimensional case goes as 
follows~\cite{fedoryuk, pham, kaminski}. Assume that the action depends on $N$
complex variables $z_i$ with $i=1,\ldots,N$.
The downward flow $F_\tau: \mathbb{C}^N\to\mathbb{C}^N$, 
$z_0\mapsto F_\tau(z_0)=z(t)$ is defined by the 
system of equations 
\beq
\frac{dz_i}{d\tau} = - \left(  \pd S{z_i}  \right)^*\quad\text{with initial condition}
\quad z_i(0) = (z_0)_i \quad \text{for $i=1,\ldots,N$.}
\label{eq:gradient-flow}
\eeq 
The role of the steepest descent path is played by a  manifold with $N$ real 
dimensions (embedded in a space of $2N$ real or $N$ complex dimensions). 
A point  belongs to this manifold -- the Lefschetz thimble $\mathcal{J}$ of the 
corresponding critical point $z_{cr}$ -- if the 
takes that point (asymptotically) to the critical point, namely 
$\lim_{\tau\to\infty}F_\tau(z) = z_\text{cr}$. Another way to visualize the thimble is 
the following: the critical point $z_\text{cr}$ is a saddle point. In the infinitesimal
neighborhood of it, there are $N$ (real) directions along which the real part of the 
action increases. Starting infinitesimally away from $z_\text{cr}$ and following the 
flow equation \eqref{eq:gradient-flow} along one of these directions defines a ray. 
The collection of all such rays is an $N$ dimensional real manifold which is the 
Lefschetz thimble associated with that critical point.  

Some intuition about the flow equations can be gained by splitting each variable $z_i =x_i + i y_i$ into its real and imaginary parts:
\bea
\frac{dx_i}{d\tau} &=& -    \frac{\partial S_R}{\partial x_i}       =     \frac{\partial S_I}{\partial y_i}   ,  \nn\\
\frac{dy_i}{d\tau} &=& -   \frac{\partial S_R}{\partial y_i}      =     -   \frac{\partial S_I}{\partial x_i} .
\eea 
The first equality shows that flow is the gradient flow for the real part of the action, $S_R$, while the second equality states that the flow is  a hamiltonian flow with the imaginary part of the action, $S_I$, playing the role of the hamiltonian which is a conserved quantity along the flow. In other words, the flow is in the direction of the fastest increase of $S_R$ and keeps $S_I$ constant.  

In general, the integration over the real variables equals the integration over a linear combination (with integer coefficients that may be zero) of the integrals over all thimbles (labeled by ``$\sigma$")  associated with critical points $z^{(\sigma)}_\text{cr}$ :
\bea
\langle \mathcal{O}[x] \rangle  
&=&
\frac {  \int \! dx\ e^{-S[x]}  \mathcal{O}[x]  }   {    \int \! dx\ e^{-S[x]}   } =  \frac{ \sum\limits_{\sigma}\  \!\!  n_\sigma \  e^{-i S_I[z^{(\sigma)}_\text{cr}] }\int_{\mathcal{J}_\sigma}\   \!\!  dz \ e^{-S_R[z]}  \mathcal{O}[z] } 
{ \sum\limits_{\sigma}\, n_\sigma  \,e^{-i S_I[z^{(\sigma)}_\text{cr}] }
\int_{\mathcal{J}_\sigma}  \! dz \, e^{-S_R[z]}    } . 
\eea 
The coefficients $n_\sigma$ are determined by how the original integration region 
(the real hypersurface $y_i=0$) intersects the unstable thimble, 
${\cal K}_\sigma$, which is defined like the thimble $\mathcal{J}_\sigma$ 
but with the reversed flow, that is the flow along which $S_R$ decreases.

The volume element in the integrals above can be defined by choosing a parametrization 
of the thimble by $N$ real variables $\eta_i$ as
\beq
dz\equiv\prod\limits_{i=1}^N dz_i = \left(\prod\limits_{j=1}^N d\eta_j\right)  \det  \underbrace{  \left( \frac{\partial z_i}{\partial \eta_j}   \right) }_{J_{ij}}\equiv d\eta \,\det J .
\eeq 
This form is not amenable for Monte Carlo computations since the Jacobian $J$ is complex. 
Let us separate the magnitude and phase of the determinant of the complex matrix $J$ and 
write $\det J = |\det J| e^{i\alpha}$. The residual phase  $e^{i\alpha}$ describes 
the inclination of the space tangent to the thimble in relation to the real hyperplane.
We have now (assuming for notational simplicity that only one thimble contributes):
\beq
\label{eq:parametrization}
\langle \mathcal{O}[x] \rangle  
=
\frac{ \int\!\! \,    d\eta \  |\!\det J|   e^{i\alpha}   \,e^{-S_R[\eta]  }  \mathcal{O}[\eta]}
{\int\!\! \,    d\eta \  |\!\det J|   e^{i\alpha}   \,e^{-S_R[\eta]  } }  = \frac{ \int\!\!  \, d\eta \     e^{i\alpha}  \, e^{-S_R[\eta]+\log( |\!\det J| )}\,  \mathcal{O}[\eta]   }
{\int\!\!   \,d\eta\   e^{i\alpha}  \, e^{-S_R[\eta] +\log(|\!\det J|) } }  =
\frac{ \langle  e^{i\alpha}    \mathcal{O}\rangle_{J}    }   {\langle e^{i\alpha}\rangle_{  J }  },
\eeq 
where the averages $\langle\cdot\rangle_J$ are defined by the effective action $S_\text{eff}\equiv S_R-\log(|\!\det J|)$. The feasibility of this form for Monte Carlo evaluations hinges on the size of the fluctuations of the residual phase $e^{i\alpha}$. This phase is very different from the phase $e^{-i S_I[x]}$ that caused the sign problem to begin with. In fact, there are simple models where $e^{-i S_I[x]}$ is rapidly oscillating (for real $x$) but $\alpha=0$
\footnote{For a trivial example, consider the one dimensional action $S(x) =(x+1000 i)^2 $.}.

\section{The algorithm}
\label{sec:algorithm}

All Monte Carlo algorithms rely on a Markov chain taking, at every step,  one point of the integration region to another point. This presents a difficulty for an integration over the thimble as the shape of the thimble is only defined via the solution of a differential equation (being a multidimensional manifold, storing the coordinates enough points on the thimble---let alone computing them!---is unfeasible). If the Markov chain is at one point of the thimble it is unclear which directions can be proposed for next step.  In other words, the tangent space to the thimble is not known locally. 

The tangent space to the thimble at the critical point is, however, relatively easy to find. In fact, near the critical point the real part of the action is given by:
\bea
S_R[z_{cr}+\Delta z] 
&\approx&
 S_R[z_{cr}] + \frac{1}{2} \Re \left(\left. \frac{\partial^2 S}{\partial z_i \partial z_j}\right|_{z_{cr}}  \Delta z_i \Delta z_j + \cdots   \right) \nn\\
 &=&
 S_R[z_{cr}]  +  \frac{1}{4} \left(  
\left. \frac{\partial^2 S}{\partial z_i \partial z_j}\right|_{z_{cr}}  \Delta z_i \Delta z_j 
 +
\left. \left( \frac{\partial^2 S}{\partial z_i \partial z_j}\right)^*\right |_{z_{cr}}  \Delta z_i^* \Delta z_j^*  +\cdots \right).
 \eea 
The tangent space consists of the directions along which $S_R$ increases. Consider the following equation:
\beq
H(z_\text{cr})^* \rho_\lambda^* = \lambda \rho_\lambda \quad\text{with $\lambda\in\Real$}\,,
\label{eq:eigens}
 \eeq 
where the hessian $H_{ij} \equiv \partial^2S/\partial z_i \partial z_j$ is 
a symmetric and in general complex matrix (thus, the matrix is not hermitian generically).  
Taken as a real vector space, there are $2N$ linearly independent ``eigenvectors" 
$\rho_\lambda$ with $2N$ real ``eigenvalues'' $\lambda$.\footnote{If only {\textit real} combinations are allowed, there are $2N$ linearly independent such vectors. If complex linear combinations are allowed only $N$ of them are independent.} 
Strictly speaking, they are not actually eigenvalues and eigenvectors of $H(z_\text{cr})$ 
due to 
the complex conjugation involved, but with a slight abuse of terminology we shall 
refer to them as such. We note first that solutions of \eq{eq:eigens} always exist with
$\lambda\in\Real$: if $(\lambda,\rho)$ is a solution, then $(e^{2i\alpha}\lambda, e^{-i\alpha}\rho)$ is also a solution. Therefore a complex $\lambda$ that satisfies \eq{eq:eigens} can always be rotated to a real number. For the existence of such solutions
let us rewrite \eq{eq:eigens} separated into its real and imaginary parts assuming 
$\lambda$ is real 
\begin{equation}
{\cal H}_\text{cr} \left( \begin{matrix} \Re \rho_\lambda \\    \Im\rho_\lambda  \end{matrix}\right) = \lambda \left( \begin{matrix} \Re\rho_\lambda \\    \Im\rho_\lambda  \end{matrix}\right)
\quad \text{with}\quad {\cal H}_\text{cr}\equiv\left(   \begin{matrix}
      H_R & -H_I \\
      -H_I & -H_R \\
   \end{matrix}\right)\Bigg|_{z=z_\text{cr}}
\,.
   \label{eq:H-matrix}
\end{equation}
The matrix ${\cal H}_\text{cr}$ is real and symmetric therefore it has $2N$ real eigenvalues 
and eigenvectors. These eigenvalues and eigenvectors are used to construct the solutions
of Eq.~\ref{eq:eigens}. We stress that they are not the eigenvalues of the complex 
matrix $H(z_\text{cr})$, 
but instead they are eigenvalues of the $2N$ dimensional real matrix ${\cal H}_\text{cr}$. 
Secondly we note that for every  $(\lambda,\rho)$ there exists $(- \lambda, i \rho)$.
\footnote{$i\rho$ is linearly independent from $\rho$ if one is allowed to make only {\it real} linear combinations of the eigenvectors.} As such, there are $N$ real positive 
eigenvalues and $N$ negative ones.
Choosing the displacements $\Delta z=\sum_{\lambda>0} c_\lambda \rho_\lambda$ 
with $c_\lambda\in\Real$ in the subspace generated by real linear combinations of 
the eigenvectors $\rho_\lambda$ with positive eigenvalues $\lambda>0$, 
increases the real part of the action:
\beq
S_R[z_{cr}+\Delta z]  \approx S_R[z_{cr}]  +\frac12\sum_{\lambda>0} c_\lambda \lambda 
\Vert\rho_\lambda \Vert^2 +\cdots > S_R[z_{cr}]  \,.
\eeq 
The collection of $N$ complex vectors $\rho_\lambda$ with $\lambda>0$ spans the tangent 
space to the thimble and it can be computed from \eq{eq:H-matrix} in a straightforward
fashion. This method of constructing the tangent space only works at the critical point.  For non-critical points, where $\partial S/\partial z_i \neq 0$, the argument does not apply.

\begin{figure}[t]
\includegraphics[width=0.45\textwidth]{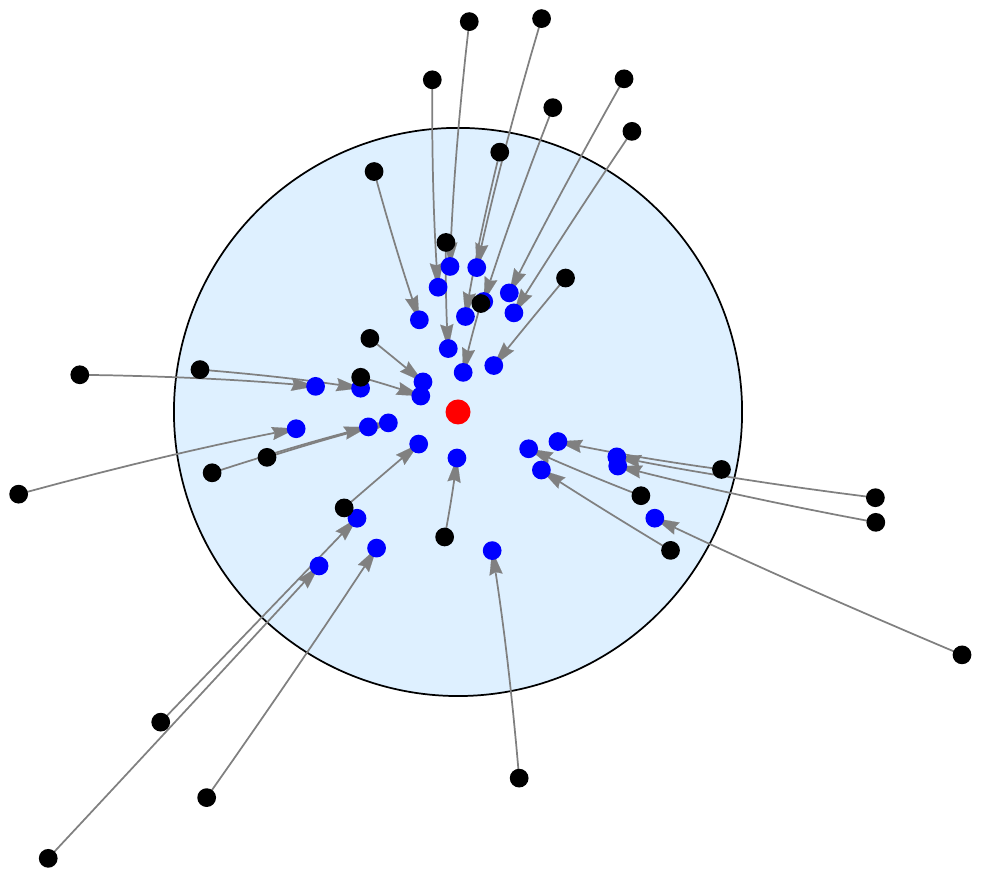}
\caption{ Schematic representation of the mapping between points $z_f$ on the thimble (black points) and their images $z_n$ (blue points) 
 in the gaussian region (shown as the light blue disk). The novelty of our algorithm is that by sampling the distribution of the blue points we can compute the integral over the whole thimble via the contraction map.}
 \label{fig:contraction_example}
\end{figure}
 
All algorithms put forward up to now rely on this fact in order to generate a Markov chain along the thimble. Every Monte Carlo update involves flowing ``downhill" to a point close to the critical point, changing the directions along the tangent plane (that is known in that region by the observations above) and flowing back near the previous point in the thimble. These methods are further complicated by the fact that the downhill flow towards the critical point is unstable, as evidenced by the presence of $N$ negative eigenvalues.

The method we use is to parametrize a generic (far) point $z_f$ on the thimble by a 
point  $z_n$ near the critical point obtained by flowing $z_f$ downhill by a fixed 
amount of time $\tau=T$, $z_n=F_T(z_f)$.
If the flow time $T$ is large enough, the relevant region of the thimble that one would like to sample (the one with significant statistical weight) is mapped into a a small region near the critical point. In this small region, which we will refer to from now on as the ``gaussian region",  the tangent space is  (approximately) found by the diagonalization of the hessian and updates can be made while staying (nearly) on the thimble. In order to illustrate this mapping we consider a two dimensional integral.
In this case the thimble is two-dimensional and by projecting on the real parts of coordinates we can depict sampled points $z_f$ as shown in \fig{fig:contraction_example}. The arrows connect the points on the thimble ($z_f$) and their respective images in the gaussian region ($z_n$). With this contraction map the sampling of points can be easily done in the gaussian region where the tangent space is computed in a straightforward fashion. The sampling statistics is determined by the images of these points in the far region therefore the integral over the full thimble is simply obtained by flowing these points by a fixed flow time. This way, as opposed to other algorithms, we avoid flowing in the unstable downward direction which causes great problems.

The near point $z_n$ plays the role of the real parameter $\eta$ in \eq{eq:parametrization}.
In fact this parametrization amounts to a change of variable in the integral:
\beq
\int dz_f \, e^{-S_R[z_f]} {\cal O}[z_f] = \int dz_n \,\det J\, e^{-S_R[z_f(z_n)]}
{\cal O}[z_f(z_n)] \quad\text{with}\quad J_{ij} = \pd{(z_f)_i}{(z_n)_j}\,.
\eeq 
Our algorithm will generate a set of points $z_n$ with the probability distribution
controlled by the Boltzmann factor:
\beq
\label{eq:prob}
P(z_n) \propto e^{-S_R[z_f(z_n)]} |\!\det J| \,.
\eeq
Note that the Jacobian $J$ corresponds to the map $z_n\mapsto z_f=\overline{F}_T(z_n)$,
that is the upward flow map for time T. The determinant of the Jacobian measures the 
inverse ratio of a volume element (an infinitesimal parallelepiped) at the near point
and the volume of its image at the far point. To compute this, we setup an infinitesimal 
parallelepiped $P_n$ at the near point $z_n$ spanning its tangent space and transport it 
using the upward flow to $z_f$ to get a parallelepiped $P_f$.
To compute the image of a vector in the tangent space, let us consider a pair of infinitesimally close points  $z$ and $z'$. 
By transporting both of them by a time step $dt$ using the upward flow we find that:
\beq
 z_i'(t+dt) - z_i(t+dt) 
 \approx
  z_i'(t) - z_i(t) +\left[  \left. \left(  \frac{\partial S}{\partial z_i}\right)^* \right|_{z'}  - \left.\left(  \frac{\partial S}{\partial z_i}\right)^* \right|_{z}  \right] dt
 \approx
  z_i'(t) - z_i(t) +\left[   \left.\left(  \frac{\partial^2S}{\partial z_i\partial z_j}\right)\right|_z (z'-z)_j \right]^* dt,
\eeq 
which shows that tangent vectors $v$ are transported by the flow according to the equation
\beq\label{eq:vector_transport}
\frac{d v_i}{d\tau} =  \left(  \frac{\partial^2 S}{\partial z_i\partial z_j} v_j  \right)^*
\quad\text{that is}\quad \dd v\tau = [H(z) v]^* \,.
\eeq 
Coupled with the differential equations for the upward flow, 
\beq
\frac{dz_i}{d\tau} = \left(  \pd S{z_i}  \right)^*\quad\text{with initial condition}
\quad z_i(0) = (z_n)_i \quad \text{for $i=1,\ldots,N$.}
\label{eq:upflow}
\eeq 
this equation allows us to
map the tangent space at $z_n$ to the tangent space at $z_f$.

To construct the parallelepiped $P_n$ we take advantage of the fact that the tangent
space in the gaussian region is well approximated by the span of the positive eigenvectors
of ${\cal H}_\text{cr}$. We set:
\beq
(P_n)_{ij} = (\rho_{\lambda_j})_i \quad\text{with $\lambda_j>0$ for $i=1,\ldots,N$,}
\eeq
and get $P_f=P(\tau=T)$ by integrating the upward flow equation
\beq
\dd P\tau = [H(z) P]^* \quad\text{with initial condition}\quad P(0)=P_n \,,
\label{eq:P}
\eeq
for time $T$. The determinant of $J$ is then
\beq
\det J = \det P_f / \det P_n \,.
\eeq
Before we describe the algorithm, we note that since $P_n$ is the same for all $z_n$'s
its contribution drops out when considering averages over one thimble. We can then
use $\det P_f$ instead of $\det J$ in the expression for the effective action. Furthermore,
we note that $P_n$ can be any set of $N$ linearly independent vectors that span the 
tangent space at $z_\text{cr}$, which means that $P_n$ is the same with the one we 
defined here up to a multiplication with an $N\times N$ non-singular real matrix.
In particular, if the eigenvectors of $\rho_\lambda$ are all real, we can set 
$P_n=\openone$.
 
In order to sample the thimble we use a simple Metropolis algorithm based on the representation of the expectation values given by \eq{eq:prob} and generate ``near''
points with the distribution $P(z_n)\propto\exp(-S_\text{eff}[z_n])$.
We use the following steps:
\begin{enumerate}
\item begin with the system at the critical point: $z_n=z_\text{cr}$.
\item pick a random vector $\delta\in\Real^N$ and make a proposal 
$z_n'=z_n + \sum_j \delta_j  \rho_{\lambda_j}$, where ${\lambda_j}>0$.
In order to insure detailed balance, the probability distribution for $\delta$ has
to satisfy $P(\delta) = P(-\delta)$.
\item  compute $z_f'$ and $\det J[z_n']$ by solving \eq{eq:upflow} and  \eq{eq:P} with initial conditions $z(0)=\tilde z_{n}'$ and $P(0)=P_n$.
\item  compute the effective action $S_\text{eff}[z_n'] = S[z_f']-\log(|\!\det J[z_n']|)$.
\item  accept/reject the proposal with probability $\min\{1, \exp(-S_\text{eff}[z_n'] + S_\text{eff}[z_n])\} $ and set $z_n$ to  $z_n'$.
\item go back to 2 and repeat the process.
\end{enumerate}
The above algorithm generates a series of triplets $(z_n, z_f, J)_k$ 
which are then used to estimate the observables average:
\beq
\left\langle {\cal O} \right\rangle \approx \frac{\sum_k {\cal O}[(z_f)_k] e^{i\arg J_k}}
{\sum_k e^{i\arg J_k}} \,. 
\eeq
Note that only the far points $z_f$ and the phase of the Jacobian are used in the
calculation for the observables.
The algorithm proposed here is simple to implement and the only delicate part is 
the integration of the downward flow, which is, as we stressed earlier, numerically stable.
We use an fifth order adaptive integrator based on Runge-Kutta method~\cite{Cash:1990:VOR:79505.79507}. The choice of the probability distribution for the random displacements,
$P(\delta)$ in step 2, is important to insure that the algorithm is efficient. We will
discuss a choice appropriate for this model in Section~\ref{sec:results}. The
results of the simulation will depend on the amount of time $T$ we integrate the 
downward flow. The exact result is recovered in the limit $T\to\infty$. In practice
this parameter will either be chosen large enough so that the results are sufficiently
close to their exact values or we can use an extrapolation in $T$.

Before we conclude, we note that a Metropolis based algorithm was 
used to investigate a simple system using Lefschetz thimble approach~\cite{Mukherjee:2013aga}.
Our proposal differs in several important details. In particular, in our work the
Jacobian is included in the effective action used for updating, whereas in the work
cited this is accounted for in the calculation of the observable as an additional
reweighting factor. For systems where the thimble is not well approximated by the
gaussian thimble, the Jacobian will fluctuate wildly and the reweighting fails.

\section{The model}
\label{sec:model}

We will illustrate our algorithm using a 0+1 dimensional version of the Thirring model at finite density which can be solved analytically and compare the Monte Carlo results to the analytical ones. This model suffers from sign problem and have been used as a toy model for testing ideas such as complex Langevin dynamics \cite{Aarts:2008rr,Pawlowski:2014ada,Pawlowski:2013pje} and hybrid Monte Carlo on Lefschetz thimbles \cite{Fujii:2015vha,Fujii:2015bua}.  The model is fermionic system with a quartic interaction and has the following continuum Euclidean Lagrangian 
\begin{equation}\label{eq:S-fermions}
L_{Th.}= \bar \chi \left (\gamma^0 {d \over dt} + m +\mu \gamma^0\right) \chi+ {g^2 \over 2}\left(\bar \chi \gamma^0 \chi \right)^2 \,,
\end{equation}
where $\chi$ is a two component spinor and $\gamma^0$ is a Pauli matrix. The interaction term is simply the 0+1 dimensional analog of the current-current interaction $(\bar\chi\gamma^\mu\chi)(\bar\chi\gamma_\mu\chi)$ of the original Thirring model. This quartic interaction term can be can be eliminated via a Hubbard-Stratonovich transformation leading to the effective Lagrangian
\begin{equation}
L= \bar \chi \left (\gamma^0 {d \over dt} +i \gamma^0 z + m +\mu \gamma^0\right) \chi+ {1\over 2 g^2} z^2 \,,
\end{equation}
where the auxiliary field $z$ is reminiscent of a one component gauge field. After integrating out fermions we arrive at the expression for the partition function
\begin{equation}
Z=\int {\cal D} z\,\det\left(\gamma^0 {d \over dt} + i \gamma^0 z +  m +\mu \gamma^0 \right) e^{- {1\over 2 g^2} \int dt  z^2 }\,.
\end{equation}
Above the Euclidean time direction is finite, with a length inversely proportional to
the temperature. The fermionic fields $\chi$ are anti-periodic and the bosonic field $z$
is periodic.
For real values of $\mu$, the determinant is complex and the model has a sign problem. 

The lattice formulation of this model with staggered fermions has the partition function \cite{Aarts:2008rr,Pawlowski:2013pje,Pawlowski:2014ada} 
\begin{equation}
Z=\left[\prod_{t=1}^{N} \int_{0}^{2\pi} {d z_t\over 2\pi}\right] \det K e^{-{1\over 2g^2} \sum_{t=1}^N (1-\cos z_t)}\equiv\left[\prod_{t=1}^{N} \int_{0}^{2\pi} {d z_t\over 2\pi}\right] e^{-S[z]}, \label{eq:Z-lattice}
\end{equation}
where the effective action and the explicit form of the discretized Dirac matrix are
\begin{eqnarray}\label{eq:S-bosons}
S[z]&=&-{1\over 2\hat g^2} \sum_{t=1}^N (1-\cos z_t)+\log\det K \,,\label{eq:S} \\
K_{t,t^\prime}&=&{1\over 2}\left(e^{\hat\mu+i z_t}\delta_{t+1,t^\prime}-
e^{-\hat\mu-i z_{t^\prime}}\delta_{t-1,t^\prime}+
e^{-\hat\mu-iz_{t^\prime}}\delta_{t,1}\delta_{t^\prime, N} -
e^{-\hat\mu-iz_{t}}\delta_{t,N}\delta_{t^\prime, 1}\right)+\hat m\,\delta_{t,t^\prime}\,. \label{eq:K}
\end{eqnarray}
Here $N=\beta/a$ is an even number that denotes the number of lattice sites related to 
the inverse temperature of the system $\beta$, and all the dimensionful quantities, 
$m$, $g^2$, $\mu$ are converted in dimensionless units by multiplying with appropriate 
powers of the lattice spacing: $\hat m=ma$, $\hat\mu = \mu a$, $\hat g^2=g^2 a$. 
The auxiliary field $z$ in this discretization plays the role of a $U(1)$ link variable. 
The partition function, the chiral condensate, and the charge density can be calculated 
analytically in this lattice model~\cite{Pawlowski:2014ada}:
\begin{eqnarray}
Z&=&{e^{-N\alpha}\over 2^{N-1}}\left[I_1^N(\alpha)\cosh(N\hat\mu)+I_0^N(\alpha) 
\cosh(N \sinh^{-1}(\hat m))\right] \,,\\
\langle n \rangle&=&{1\over \beta}{\partial \over \partial \mu} \log Z=
{1\over N}{\partial \over \partial \hat\mu} \log Z= {I_1^N(\alpha)\sinh(N\hat\mu) \over I_1^N(\alpha)\cosh(N\hat\mu)+I_0^N(\alpha) \cosh(N \sinh^{-1}(\hat m)) } \,,\\
\langle \bar\chi\chi \rangle&=&{1\over \beta}{\partial \over \partial m} \log Z=
{1\over N}{\partial \over \partial\hat m} \log Z= {(1+\hat m^2)^{-1/2}I_0^N(\alpha)\sinh(N\sinh^{-1}(\hat m)) \over I_1^N(\alpha)\cosh(N\hat \mu)+I_0^N(\alpha) \cosh(N \sinh^{-1}(\hat m)) }\,,
\end{eqnarray} 
where $\alpha\equiv1/(2 \hat g^2)$ and $I_n(\alpha)$ denotes the modified Bessel function of the first kind. 

\subsection{Semiclassical approximation and subleading thimbles}

For analyzing our Monte Carlo results it is useful to have an estimate for the
contribution of individual thimbles to the final result, in particular the leading and
subleading thimbles. We note that for fermionic systems the analysis is more involved
due to the zeros of the determinant, which lead to divergencies in the effective
action. The thimbles that start at the critical points could run to infinity but they
can also terminate at a zero of the determinant. In this section we will focus on estimating
the contribution of the subleading thimbles to gauge the expected discrepancy between
the Monte Carlo one-thimble result and the exact result that should be exactly reproduced
only when all contributing thimbles are included in the Monte Carlo calculation.

The critical points are determined by the equation
\begin{equation}
{\partial S \over \partial z_t}={1\over 2 \hat g^2} \sin(z_t) - i {\sinh\left( N \hat\mu + i \sum_{t^\prime=1}^N z_{t^\prime} \right) \over  \cosh\left( N \hat\mu + i \sum_{t^\prime=1}^N z_{t^\prime}\right)+\cosh(N \sinh^{-1}(\hat m)) }=0 \label{eq:semiclassical}\,, 
\end{equation}  
which we solve numerically. We want to stress two points here:
First, the second term in Eq.~\ref{eq:semiclassical} is independent of $t$ which leads to 
the conclusion that all the critical points of the discretized path integral in 
Eq.~\ref{eq:Z-lattice} have the property that $\sin z_t$ is time independent, that is
$\sin z_t=\sin z_0$. If the field
was continuous this would imply that the $z_t$ is also time independent 
(that is $z_t=z_0$ for all $t$), but in a 
discretized system it is possible to have critical points where the value of $z_t$ changes 
to $\pi-z_0$. However, we expect that the leading contributions come from thimbles attached
to critical points with constant $z_t$. This assumption substantially simplifies the 
problem of finding the critical points of the action.   
Secondly, the leading contribution comes from the thimble attached the critical point 
with $\Re z_0=0$. For $\hat\mu=0$ this critical point is at $z_0=0$ but 
for nonzero chemical potential it is a pure imaginary, that is $z_0= i x$ for 
some real $x$, even though the original path integral is along the real values 
of $z_t$, namely the interval $[0,2\pi]$. This is an example of the situation where 
complex valued configurations which lie outside of the original integration 
region contribute to the semiclassical 
expansion~\cite{Basar:2015xna,Behtash:2015kna,Behtash:2015kva,Behtash:2015zha}. 
In fact in this case the complex configuration constitutes the {\em leading} contribution. 

In the semiclassical approximation we approximate the path integral by evaluating the
effective action and the observable only at the critical points.
Within the semiclassical approximation we have
\begin{equation}
\langle {\cal O}[z]\rangle \approx {\sum_\sigma n_\sigma {\cal O}[z^{(\sigma)}_\text{cr}] 
e^{-S[z^{(\sigma)}_\text{cr}]}\over \sum_\sigma n_\sigma e^{-S[z^{(\sigma)}_\text{cr}]}}
\end{equation}
where $\sigma$ labels different critical points. The coefficients $n_\sigma$ can be zero 
for some critical points. These coefficients are difficult to determine precisely given
the complicated topology of the thimbles in the ${\mathbb C}^N$ space. We compute all
critical points that have constant field, do a flow analysis in the constant field
plane and estimate the thimbles that are likely to contribute. To derive the
estimate, we assume that the values of $n_\sigma$'s are equal to $1$ for the thimbles
we believe contribute and include in our estimate the largest two sub-leading thimbles.
Given the heuristics involved in our procedure, we use these estimates only to set
the expectations for the order of magnitude of the disagreement between the Monte
Carlo and exact results.

We apply the procedure above to estimate the subleading contributions to the 
chiral condensate $\left\langle \bar{\chi}\chi \right\rangle$.
In \tab{tab:semiclassical} we give the numerical values of the contribution from the 
next-to-leading critical point for a range of parameters. The order of magnitude
for this quantity is about $1$. In the table we include the results for 
a set of parameters similar to the one used in our numerical simulations. We
set $\hat m=1$ and look at two values for the coupling $\hat g^2=1/6$ (weak coupling) and 
$\hat g^2=1/2$ (strong coupling). At low values of $\mu$ the subleading contribution
is very small, it increases for values around $\hat \mu=1$ and decreases again
as $\hat \mu$ is greater than $2$. From the table we see that for $N=2$ the discrepancy
for the weak coupling case is very small and it is unlikely to resolve it using
Monte Carlo methods. However, even for weak coupling, when the temperature is increased,
as we show in the table for $N=8$, the other thimble contributions grows and 
we expect that discrepancy to be resolved in our simulations. For strong coupling
the subleading contribution is large even at high temperature.
These values are to 
be compared with the difference between the exact result and one thimble Monte 
Carlo computation given in the bottom line of \fig{fig:n2} and the right hand side 
of \fig{fig:n8}. The order of magnitude agreement between the semiclassical estimate of 
the contribution of the subleading thimble and the difference between the exact result 
and the Monte Carlo result, $\delta\langle\bar\chi\chi\rangle$, show that 
the algorithm is correct and that the discrepancy is due to the contribution
of the neglected thimbles.

\begin{table}[t]
\begin{tabular*}{0.99\textwidth}{@{\extracolsep{\fill}}c*{15}{c}{c}@{}}
\toprule
&&\multicolumn{4}{c}{$N=2\qquad \hat g^2=1/6$}&& \multicolumn{4}{c}{$N=2\qquad \hat g^2=1/2$} &&\multicolumn{4}{c}{$N=8\qquad \hat g^2=1/6$} \\
\cmidrule{3-6}\cmidrule{8-11}\cmidrule{13-16}
  $\hat\mu$ & &  $0.6$ & $1.0$ & $1.4$ & $2.0$& &  $0.6$ & $1.0$ & $1.4$ & $2.0$& &  $0.6$ & $1.0$ & $1.4$ & $2.0$\\ 
$\delta\langle \bar\chi\chi\rangle$& & $2 \times10^{-6}$ & $4 \times10^{-4}$ &$- 4\times 10^{-4}$ &  $4 \times10^{-7}$ & & $0.05$ & $0.03$ &$0.004$ &  $-0.01$ &&  $-0.08$ & $0.004$ &$0.02$ &  $1\times10^{-8}$\\
\bottomrule
 \end{tabular*}
 \caption{Semiclassical estimates for the contribution of the sub-leading thimbles to the
 chiral condensate. The contribution is significantly smaller for smaller and larger
 values of $\hat \mu$. These estimates are for systems with mass $\hat m=1$.}
  \label{tab:semiclassical}
\end{table}

\subsection{Continuum limit}
\label{sec:contlim}

It is also illustrative to study the continuum limit of the model. The continuum limit is 
a two fermion system with four levels. By taking the limit $a\rightarrow0$ while keeping 
$\beta=N a$, $m=\hat m/a$, $g^2=\hat g^2/a$ and $\mu=\hat\mu/a$ constant we obtain
\begin{equation}
Z\rightarrow e^{-\beta(-m-g^2/4)}+e^{-\beta(-\mu+3g^2/4)}+e^{-\beta(\mu+3g^2/4)}+e^{-\beta(m-g^2/4)} \,,
\end{equation}
up to an overall normalization factor that we dropped. Further shifting the ground state energy $-m-g^2/4$ to zero we obtain the spectrum: $0,m-\mu+g^2,m+\mu+g^2, 2m$. Notably when $\mu = m +g^2$, the ground state flips from the unoccupied state to a singly occupied state, a quantum mechanical analog of a phase transition. This value of $\mu$ is where the susceptibilities peak as well.

The systems we consider in this study use rather coarse lattice spacing and it is useful
to consider a continuum trajectory that approaches the limit faster. This can be determined
by casting the discretized partition function in Eq.~\ref{eq:Z-lattice} using an
ansatz $Z=\sum_i \exp(-\hat E_i N)$, where $\hat E_i = a E_i$. We find:
\beqs
\hat E_0 &= -\log |\hat m-\sqrt{1+\hat m^2}|  -\log I_0(\alpha) + \log 2 + \alpha\,,\\
\hat E_1^{\pm} &= -\log I_1(\alpha) \pm \hat\mu + \log 2 + \alpha \,,\\
\hat E_2 &= -\log |\hat m + \sqrt{1+\hat m^2}| -\log I_0(\alpha)+ \log 2 + \alpha\,.
\eeqs
Setting the ground state energy to zero, and doing an expansion in $\hat m=ma$, 
we get:
\beqs
E_1^{\pm}&=m\pm\mu+\frac1a\log \frac{I_0(\alpha)}{I_1(\alpha)} + {\cal O}(\hat m^3) \,,\\
E_2 &= 2m + {\cal O}(\hat m^3) \,.
\eeqs
Note that 
\beq
\frac1a\log \frac{I_0(\alpha)}{I_1(\alpha)} = g^2+a g^4 + {\cal O}(\hat g^6) \,,
\eeq
and this result is indeed compatible with the continuum limit derived above as it differs
only at higher order. We will use these relations to adjust the model parameters 
as we approach the continuum limit in our simulations. We set $\hat m=ma$ and 
$\hat\mu=\mu a$ but we adjust the coupling $\hat g^2$ to 
keep the quantity $[\log I_0(\alpha)/I_1(\alpha)]/a$ constant.

\section{Results}
\label{sec:results}

\begin{figure}[t]
\includegraphics[height=0.3\textwidth]{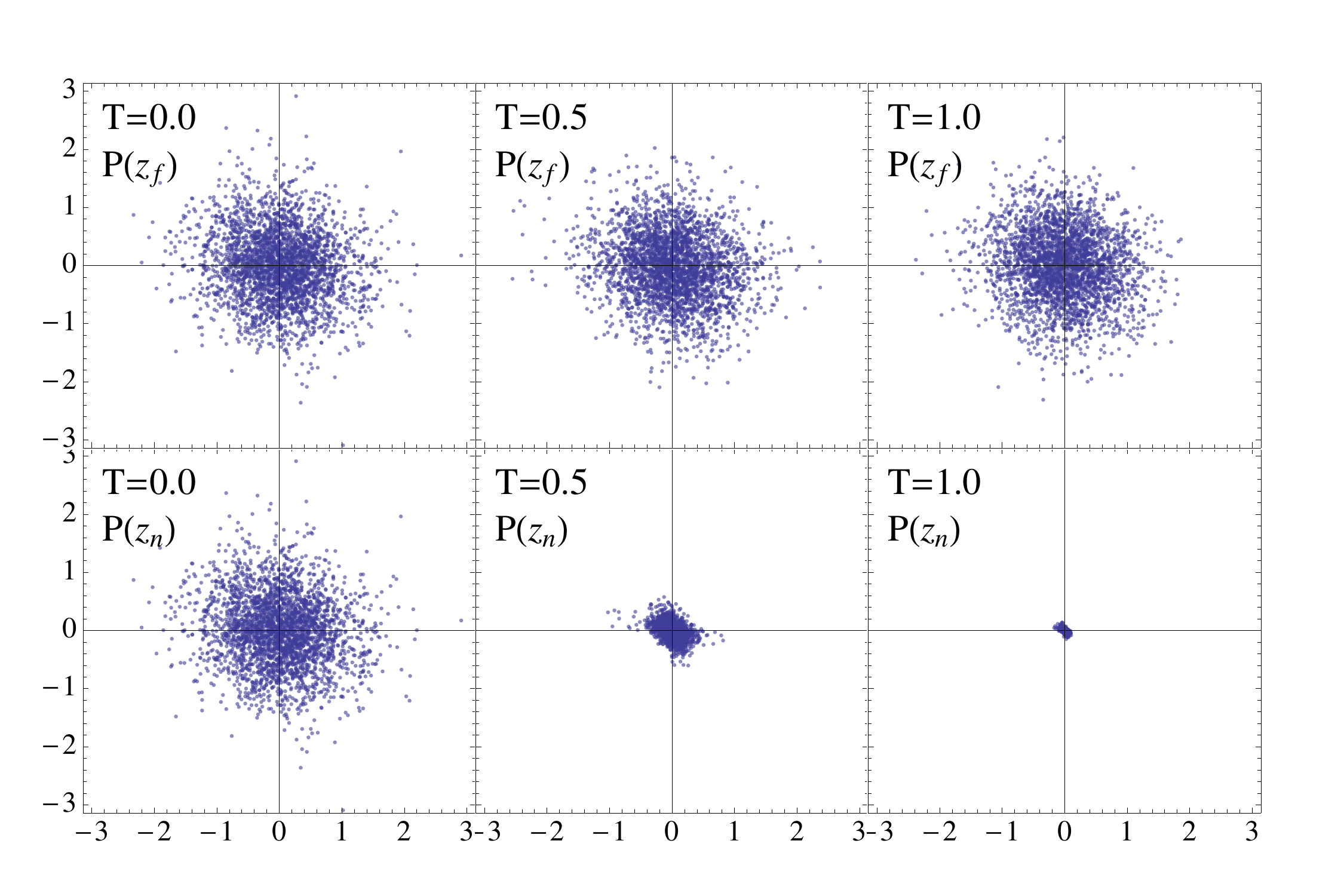}
\raisebox{-9pt}{\includegraphics[height=0.315\textwidth]{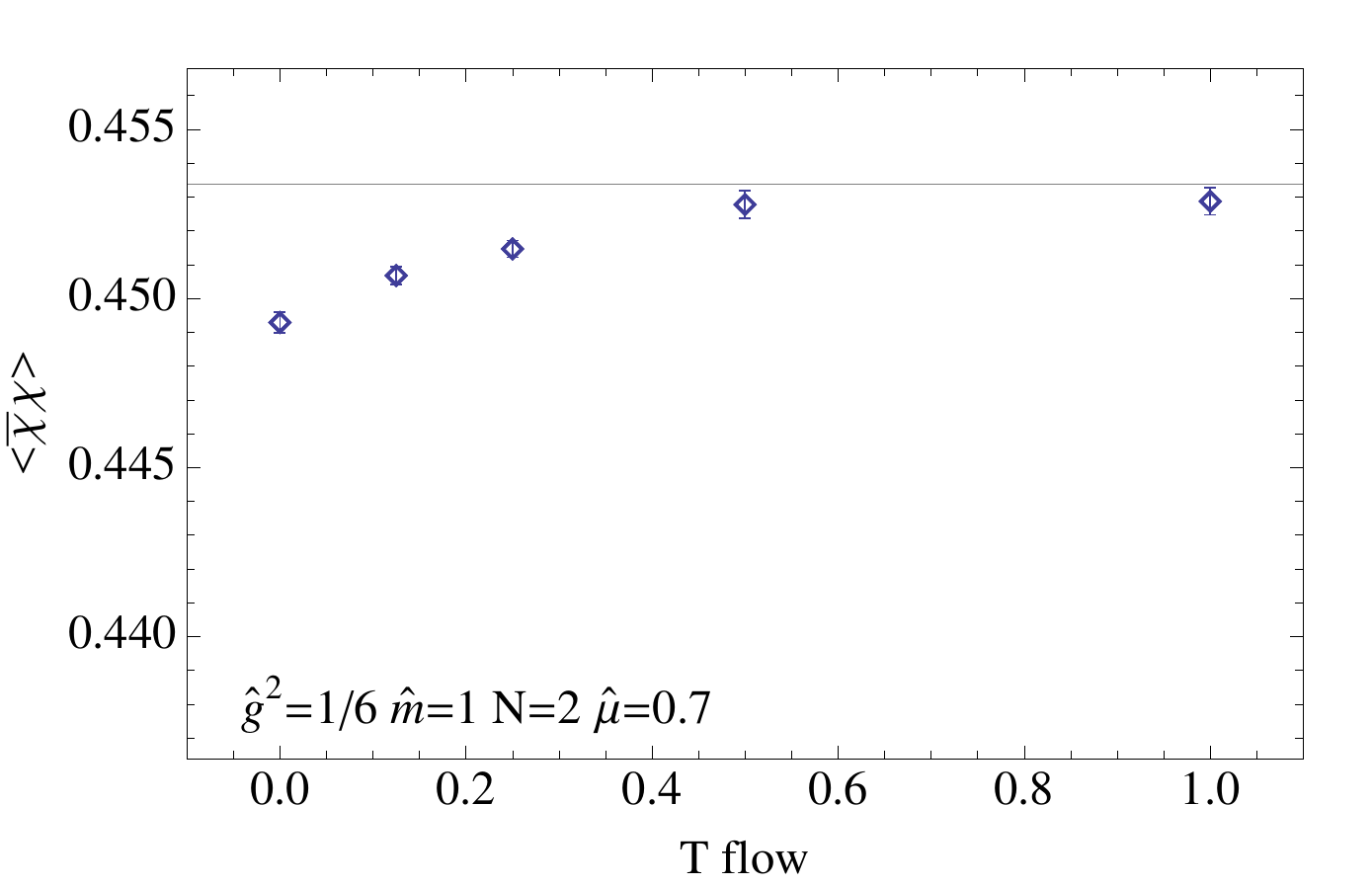}}
\caption{Left: Probability distribution for the far (top row) and near points (bottom row) 
for different flow times. We plot 3000 samples from a simulation with $N=2$, 
$\hat g^2=1/6$, $\hat m=1$, and $\hat\mu=0.7$.
Right: Chiral condensate as a function of flow time for the same parameters 
showing the convergence towards the $T\to\infty$ result. The horizontal line
indicates the exact result which is expected to be very close to the one-thimble
result for these parameters.
}
\label{fig:contraction}
\end{figure}

We performed numerical simulations of the model in \eq{eq:Z-lattice} using the algorithm 
described in Section \ref{sec:algorithm} for a variety of parameters 
$\hat m,\hat \mu,\hat g^2, N$ and flow time $T$. In this section we describe our 
main findings.

We start first by discussing some algorithmic issues. We focus first on the choice of
the flow time $T$. In the left panel of \fig{fig:contraction} we show the distribution
of the far and near points for different flow times, for the parameters set 
$\hat m=1$, $\hat\mu=0.7$, $N=2$, $\hat g^2=1/6$, and flow times $T=0, 0.5, 1$.
On the top row we show the sampled points $z_n$ (actually, a projection on the real plane).  
On the bottom row we show the image of these points $z_f$ (again, their projection on 
the real plane) connected to them by the flow. This shows that the distribution of sampled
points on the thimble is almost independent of the flow even when their corresponding 
distribution of points in the gaussian region is more and more concentrated around 
the critical point (as $T\to\infty$). These graphs also indicate how 
much flow is necessary in order to have the near points $z_n$ in the gaussian region. 
To be more quantitative, we study the dependence of the chiral condensate as a function
of the flow time. The exact result are obtained only in the $T\rightarrow\infty$ limit. 
In the right panel of \fig{fig:contraction} we show the chiral condensate as a function
of the flow time for the same parameters as in the left figure. It is clear that already
for $T=0.5$ the infinite flow time limit is reached (at the level of the error bars). 
For the calculations shown in this paper a flow time of $T\leq 3$ was always enough 
for our purposes. 

\begin{figure}[t]
\raisebox{0.36cm}{\includegraphics[height=3.95cm]{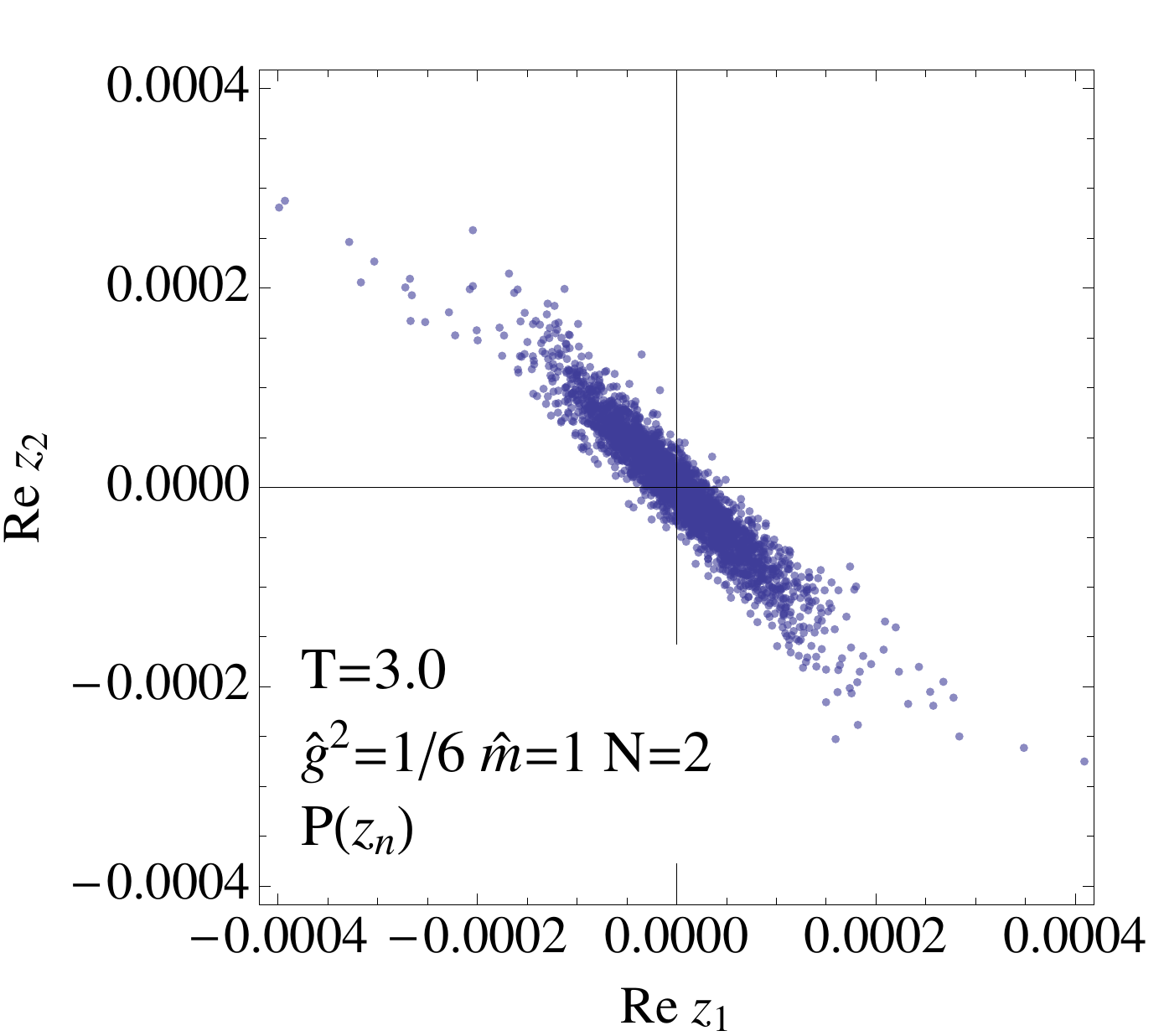}}\kern0.2cm
\includegraphics[height=4.8cm]{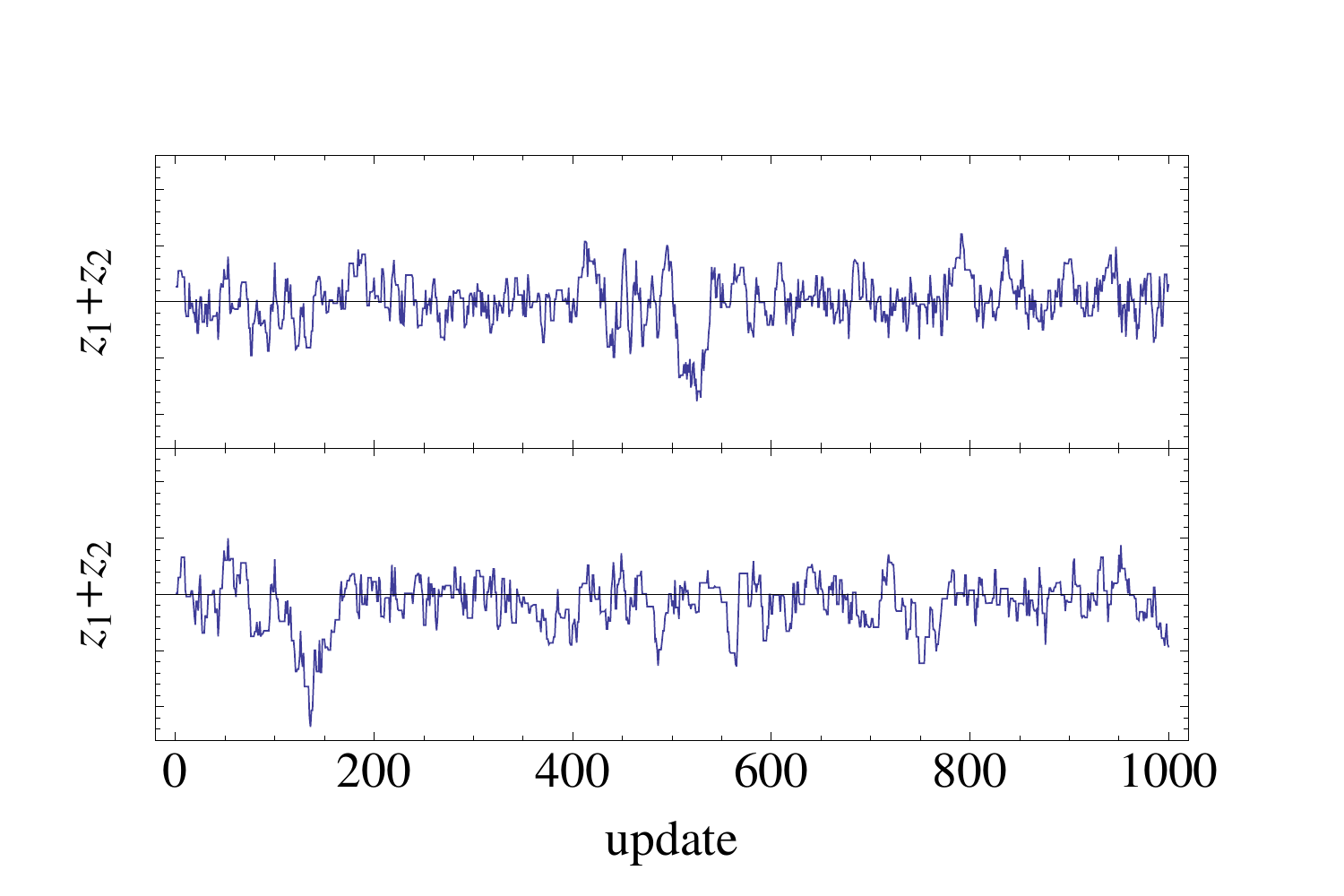}\kern-0.6cm
\includegraphics[height=4.8cm]{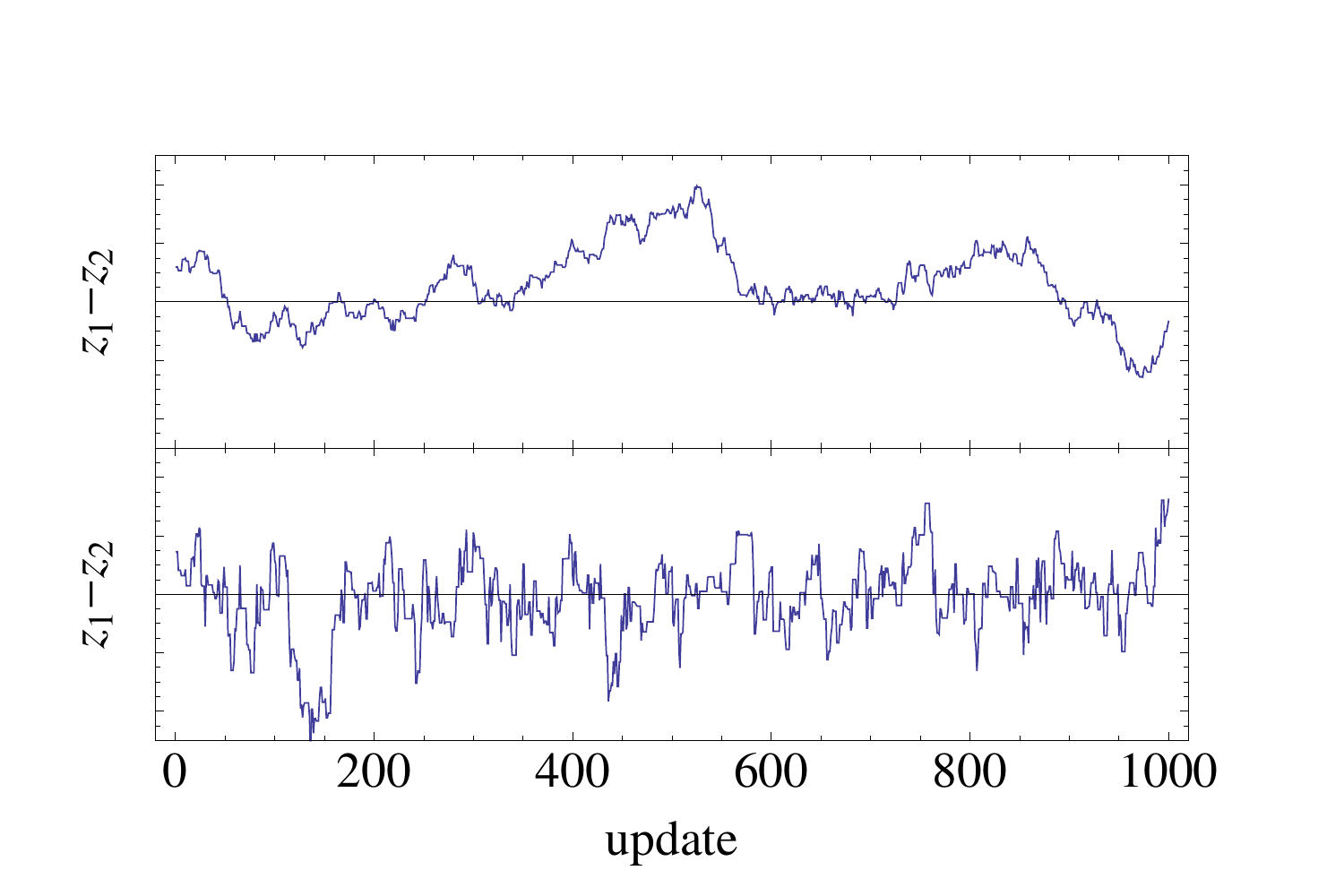}
\caption{Left:~Near points distribution, $P(z_n)$, for a flow time $T=3$ showing 
the anisotropy of their distribution. The simulation uses the same parameters as in Fig.~\ref{fig:contraction}. Middle:~Simulation time evolution for the sum
of the real parts $\Re(z_1 + z_2)$. Top is the simulation using an isotropic proposal
and bottom anisotropic. The step-sizes were adjusted to get the same acceptance rate. 
Right:~Plot of the difference $\Re(z_1 - z_2)$ which corresponds to the
elongated direction in the left panel. Note that for isotropic proposals (top) the
autocorrelation time is much longer.}
\label{fig:anisotropic}
\end{figure}

When we discussed the algorithm
in Section~\ref{sec:algorithm} we mentioned that the choice of the probability 
distribution for the proposal step $P(\delta)$ is important to insure the efficiency
of the update. To explain our choice we need to discuss the properties of the map
for large flow times. The left panel of \fig{fig:anisotropic} shows an important feature 
of the map between $z_f$ and $z_n$. Even when the points $z_f$ are distributed more or less 
isotropically their image $z_n$ can be very anisotropic.  This is due to the fact that 
the downward flow $F_T$ that maps $z_f$ into $z_n$ in the gaussian region is essentially 
a compression in the different eigendirections $\rho_\lambda$ by an amount 
$e^{-\lambda T}$. Even modest 
differences between eigenvalues will, at large $T$ produce a very anisotropic flow.
Since our algorithm samples the distribution $P(z_n)$ which is anisotropic, a isotropic 
proposal that has a good acceptance rate will be controlled by the narrowest direction
(corresponding to the largest eigenvalue $\lambda$), but it will then take a long time
to sample the ``long'' directions. In the top row of the middle and right panels of 
\fig{fig:anisotropic} we show the simulation time evolution for the eigendirection relevant
for $N=2$ system: $z_1\pm z_2$. We tune the algorithm for an acceptance rate of about 50\%
and we find that the narrow direction, $z_1+z_2$, is well sampled, whereas the long one,
$z_1-z_2$, has a large autocorrelation time.

For this reason we chose our Metropolis update proposals also anisotropically. The 
proposals in the direction $\rho_\lambda$ are proportional to the factor 
$e^{-\lambda T}$. For the model in \eq{eq:S-bosons} the eigenvalues of 
$H(z_\text{cr})$ are readily obtained because the Hessian at the critical point 
has a particularly simple structure. Remember that the critical point is purely imaginary
and constant in time $(z_\text{cr})_t = i \zeta$. The value of $\zeta$ is determined 
numerically by solving Eq.~\ref{eq:semiclassical}. All off-diagonal elements 
of the Hessian are equal to
\beq
H_{12} = \frac{\cosh[N(\hat\mu - \zeta)]}{2^{N-1} \det K(\zeta)} -
\left(\frac{\sinh[N(\hat\mu-\zeta)]}{2^{N-1}\det K(\zeta)}\right)^2 \,,
\eeq
and the diagonal elements are
\beq
H_{11} = \alpha \cosh(\zeta) + H_{12} \,.
\eeq
The eigenvalues are then $\lambda_\text{const}=H_{11}+(N-1) H_{12}$ 
(corresponding to the eigenvector with all components equal) and 
$\lambda_\text{other}=H_{11}-H_{12}$, with a $N-1$ degeneracy (corresponding to the 
remaining eigenvectors). Furthermore, note that the hessian $H(z_\text{cr})$ is real
and then the eigenvectors are purely real. Thus we do not have to solve the eigenvectors
explicitly either for the purpose of updating $z_n$, nor for computing the parallelepiped
$P_n$ required to determine $\det J$. 

Our update procedure is then $z_n'=z_n + \sum_\lambda \delta_\lambda e^{-\lambda T} \rho_\lambda$, with $P(\delta_\lambda)$ an uniform distribution in the interval 
$[-\epsilon,\epsilon]$, independent of $\lambda$. For most of our runs we choose $\epsilon=1$,
as we find that this choice produces good acceptance rates. We note that with this proposal,
the acceptance rates were almost independent of the flow time $T$. For simulations with
large number of time slices $N\ge32$ we had to reduce $\epsilon$ to about $0.1$ to get
good acceptance rates. This proposal method ensures 
that the thermalization of the average value of the field over time slices (corresponding
to $\rho_\text{const}$) is thermalized on the same time scale as the other directions 
in field space. As an example we show the results of anisotropic proposals in
the bottom row of the middle and right panels of \fig{fig:anisotropic}. In the right
panel we can clearly see that the autocorrelation time is much smaller when using
anisotropic proposals.

\begin{figure}[h]
\includegraphics[width=0.45\textwidth]{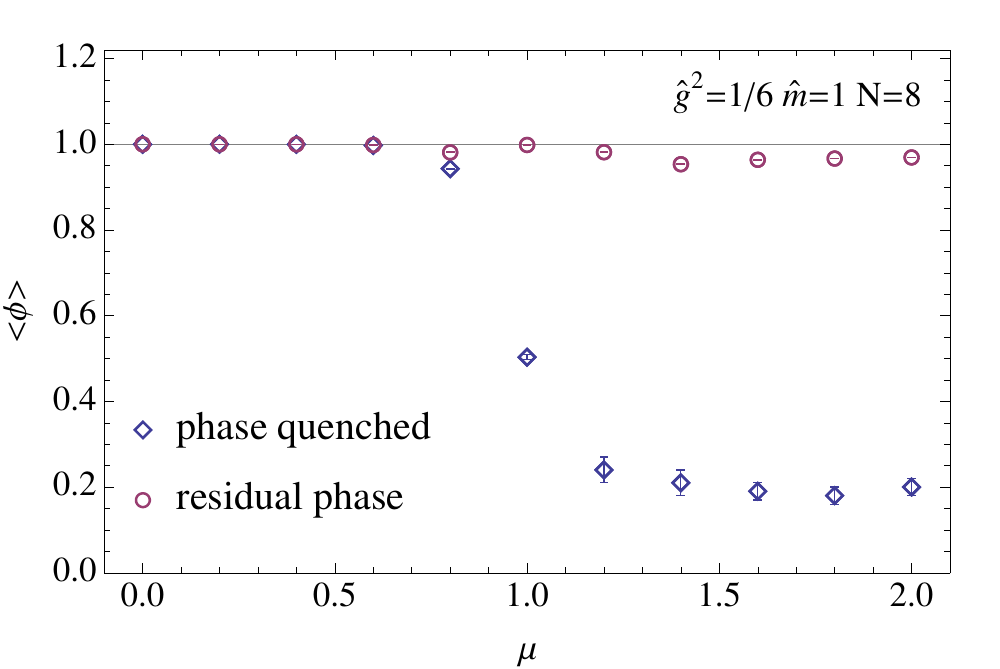}
\includegraphics[width=0.45\textwidth]{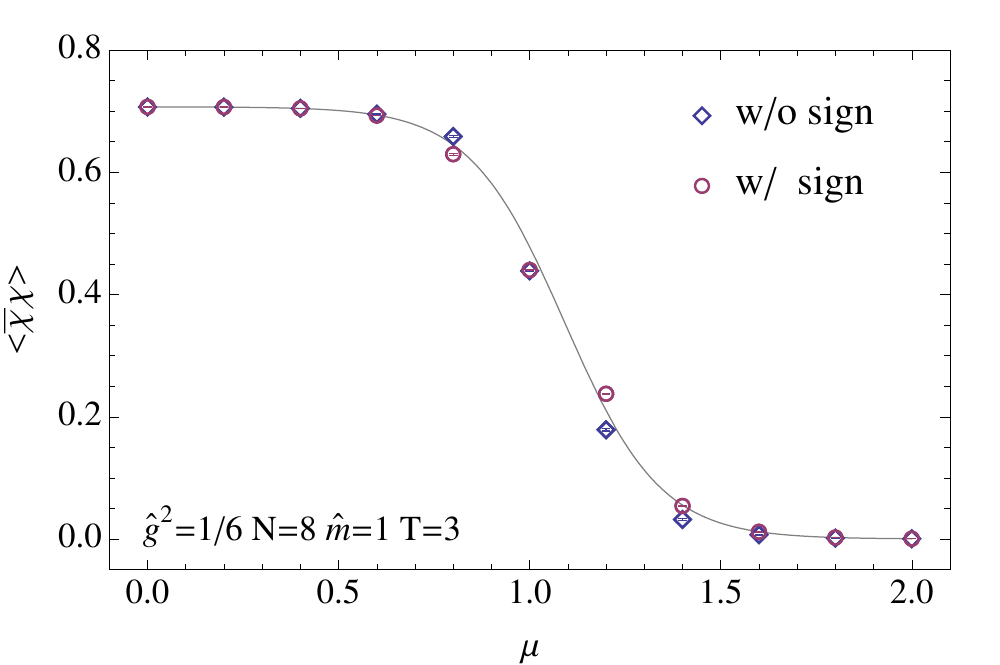}
\caption{Left: Phase average in for real phase quenched simulations and the average
of the residual phase for single thimble simulations. Right: the chiral condensate
with and without the residual phase folded in. The solid line represents the exact
result that includes contribution from other thimbles.}
\label{fig:phase}
\end{figure}

The final algorithmic issue we will address here is connected to the residual phase.
If we were to carry out the simulations using the original integration path with 
$z\in\Real^N$, we would have to do phase quenched simulation and introduce the phase
in the observable. For large $\hat\mu$ and small temperatures the average of the
phase becomes very small leading to the notorious sign problem. When integrating the
field over the thimble a similar procedure is applied to separate the complex phase
of the Jacobian. This residual phase has much smaller fluctuations and it does not
create any reweighting problems. To show this in the left panel of \fig{fig:phase}
we plot both the phase quenched average and the average of the residual phase for
a low temperature, weak coupling system. Note that as $\hat\mu$ grows bigger than $1$
the phase quench average drops dramatically, whereas the residual phase average barely
changes. However, the fluctuations of the residual phase are important and should not
be neglected: in the right panel of \fig{fig:phase} we plot the value of the chiral
condensate for the same system with the residual phase included and compare it with the
average of the observable when the phase is neglected. We can see that in the transition
region $\hat\mu\sim 1$ the difference between the two averages is noticeable.

\begin{figure}[h]
\includegraphics[width=0.45\textwidth]{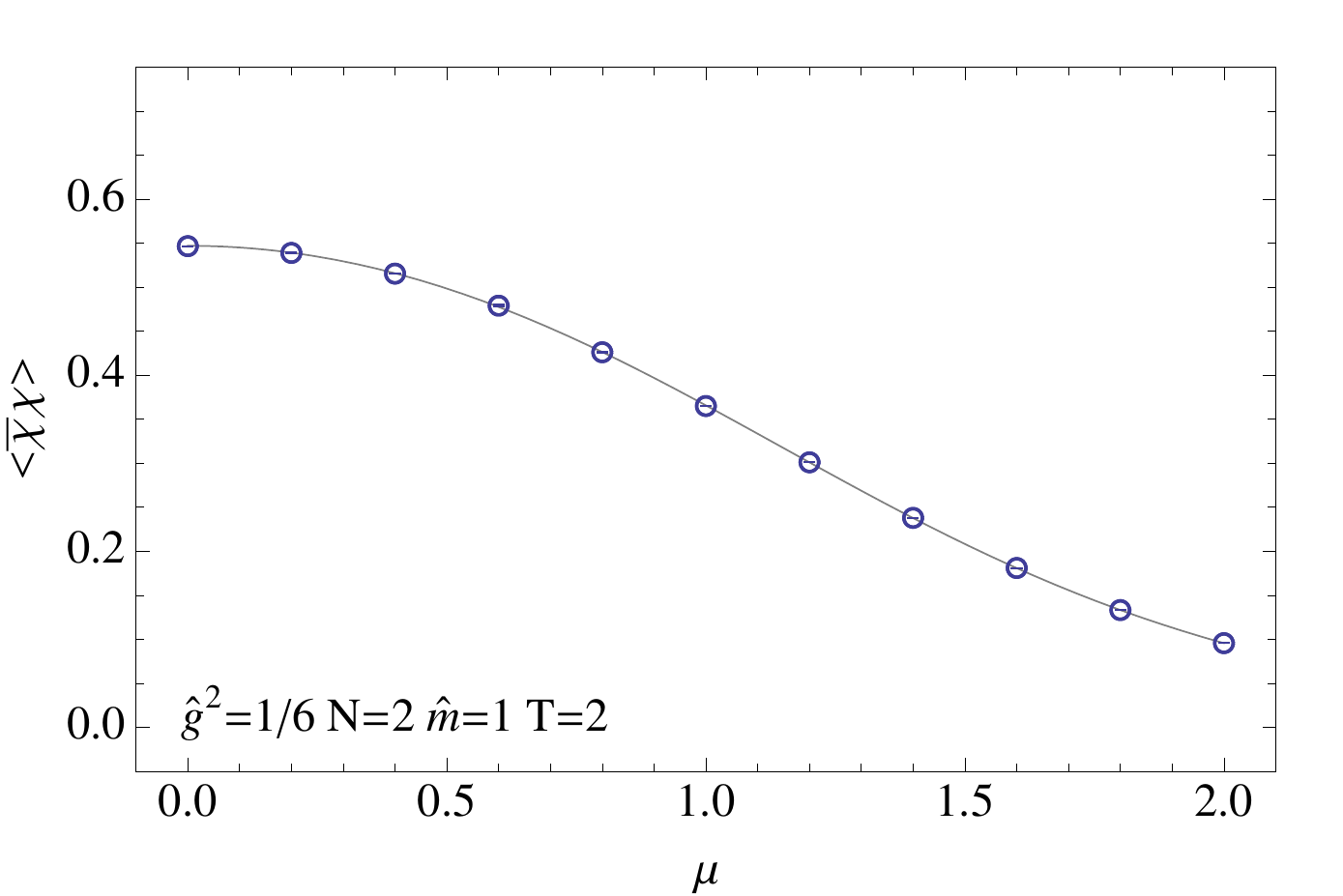}
\includegraphics[width=0.45\textwidth]{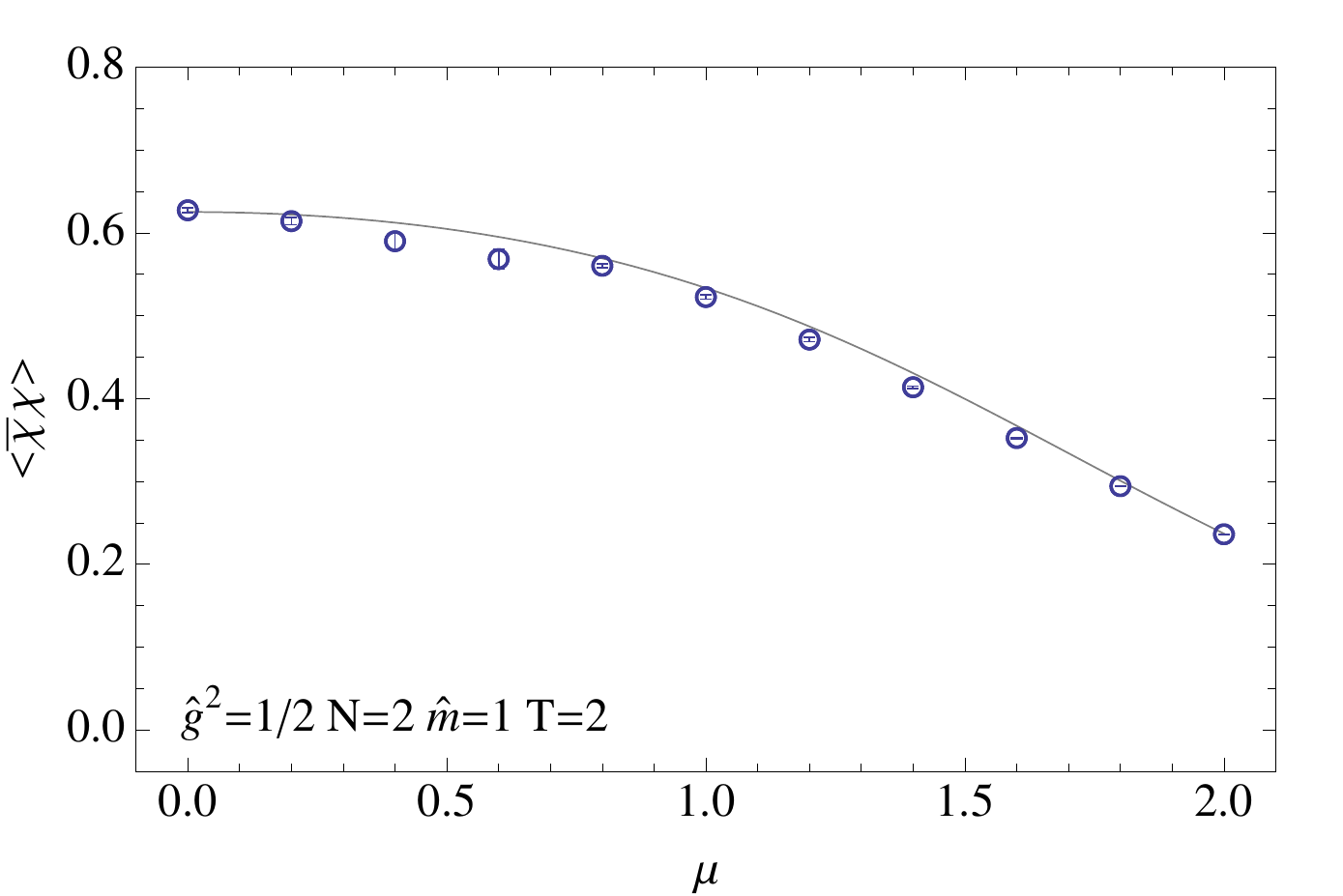}
\includegraphics[width=0.45\textwidth]{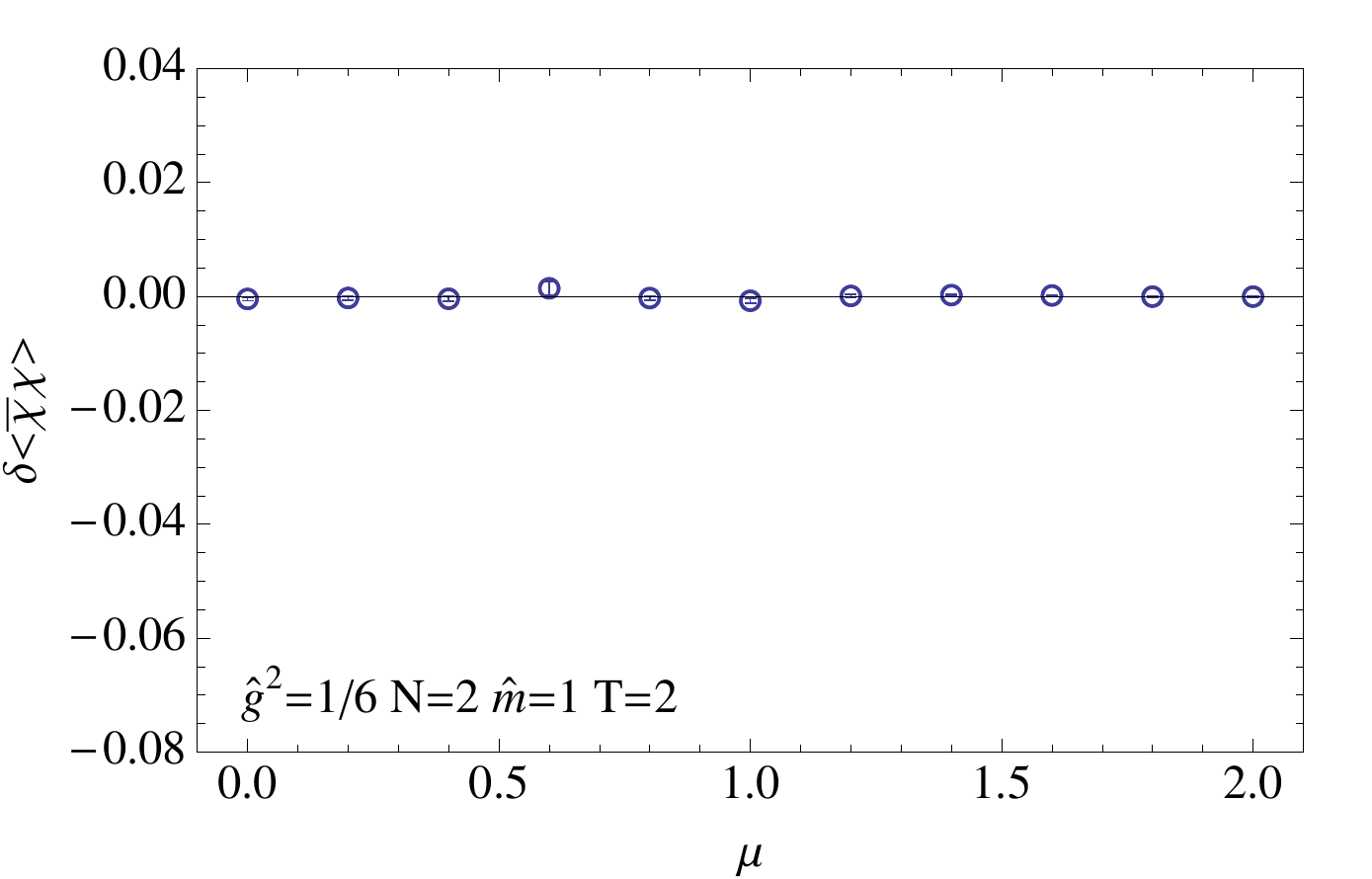}
\includegraphics[width=0.45\textwidth]{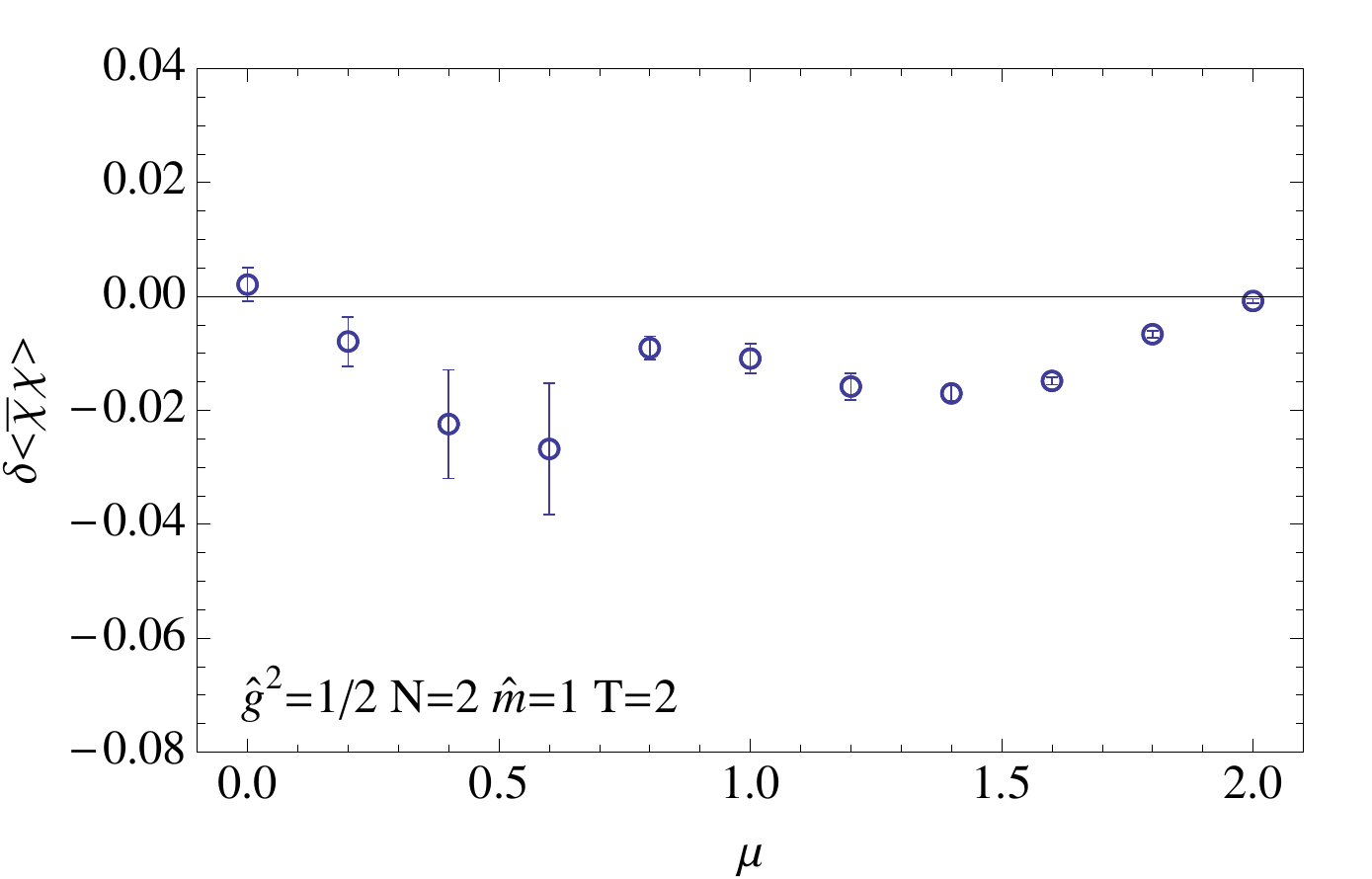}
\caption{Condensate as a function of chemical potential $\mu$ for the parameters for weak coupling (left) and strong coupling (right). In the top plots the solid lines indicate the exact result. The bottom plots indicate the difference between Monte Carlo and exact results. No discrepancy with the exact result is seen in the weak coupling case but a small statistically-significant difference is seen in the strong coupling case. }
\label{fig:n2}
\end{figure}

We now discuss the results of the Monte Carlo simulations for two sets of parameters:
weak coupling, $\hat g^2=1/6$, and strong coupling, $\hat g^2=1/2$. The mass is fixed
at $\hat m=1$. For each set of parameters we thermalized the system using 1,000 updates
and collected 10,000 samples separated by 10 updates. The error bars were estimated
using binned jackknife method using bins of size 1,000.
The results for the condensate weak coupling and $N=2$ are presented as a function of 
$\mu$ in the left panel of \fig{fig:n2}. The main feature to notice is the agreement 
between our results and the exact one. This occurs even when $\mu$ is large where the
phase quenched simulations have large phase fluctuations. This agreement is expected since 
the estimates 
for the size of the contribution from other thimbles (see table \tab{tab:semiclassical}) 
is very small, smaller than our already small error bars.
The results for strong coupling (right panel of \fig{fig:n2}) show a small but 
statistically significant difference from the exact result. This is also expected 
as table \tab{tab:semiclassical} show that the estimate of the contributions of other 
thimbles are of the same order. 

\begin{figure}[h]
\includegraphics[width=0.45\textwidth]{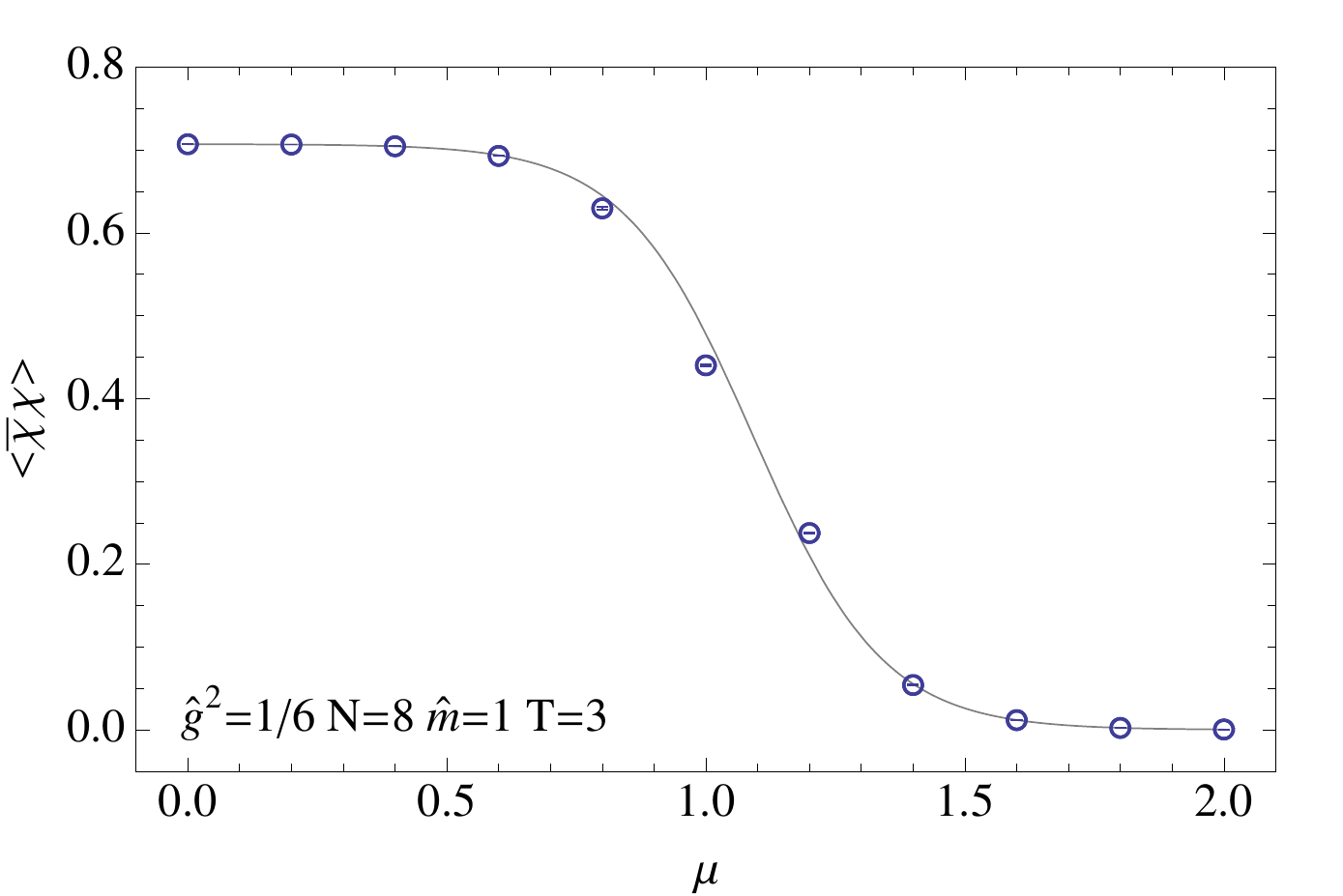}
\includegraphics[width=0.45\textwidth]{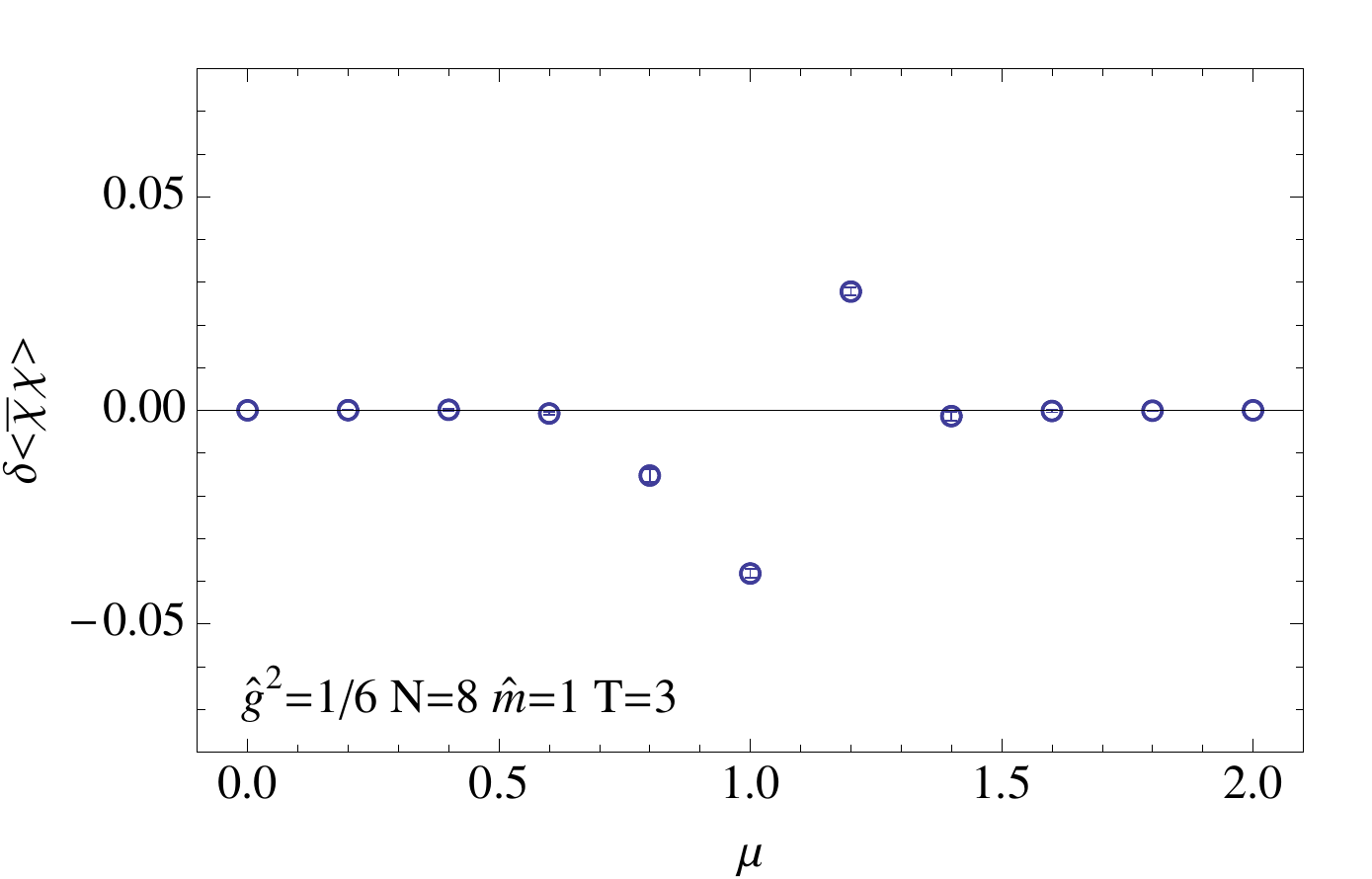}
\caption{Condensate as a function of chemical potential in the high-temperature case. 
Despite the weak coupling a discrepancy with the exact results is visible.}
\label{fig:n8}
\end{figure}

Even at small coupling the contribution of other thimbles becomes important at low 
temperatures. An example we use the small coupling parameters and set $N=8$, which corresponds
to a temperature 4 times lower than in the $N=2$ case. The results are shown in 
\fig{fig:n8} where, again, a small but statistically significant discrepancy with the 
exact result is seen. Once more, the magnitude of these deviations are comparable to 
the semiclassical estimates in \tab{tab:semiclassical}.

\begin{figure}[h]
\includegraphics[width=0.45\textwidth]{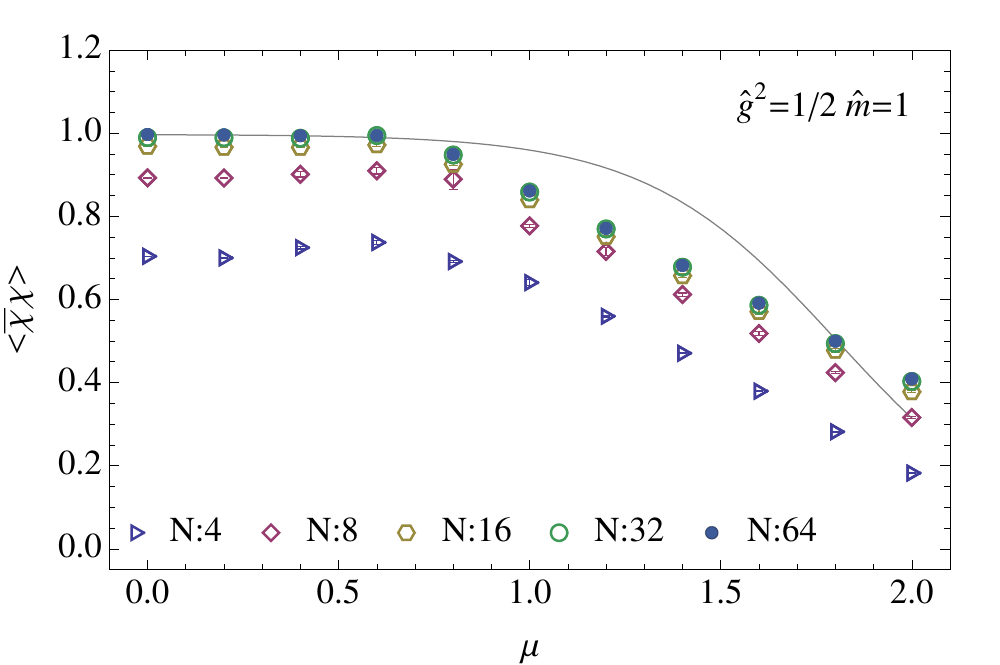}
\includegraphics[width=0.45\textwidth]{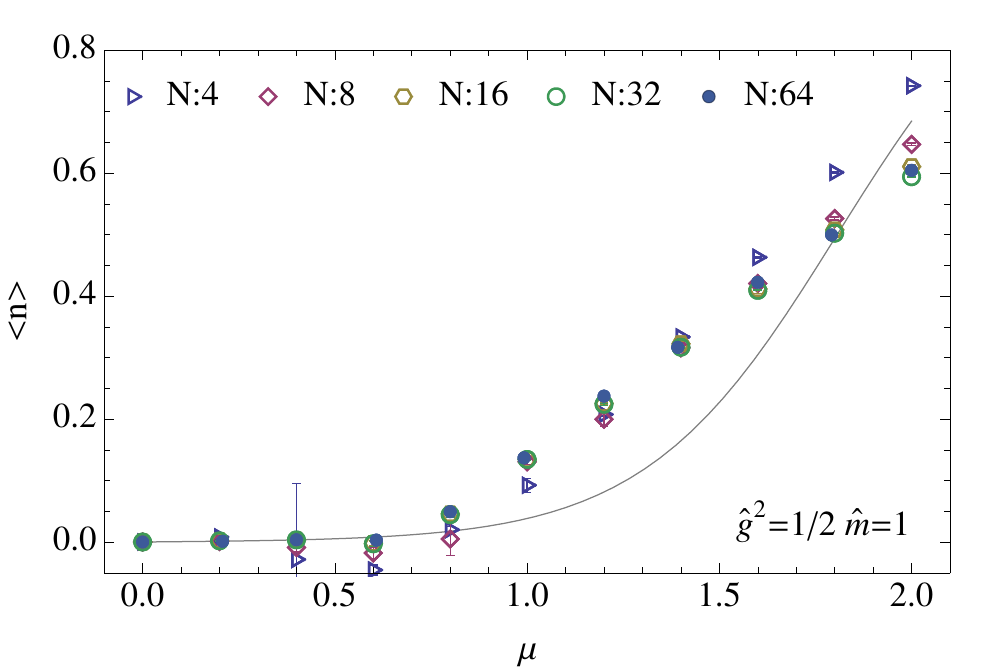}
\caption{Condensate (left) and particle number (right) as a function of chemical potential $\mu$ in the continuum limit. The solid line indicate the exact, continuum results.}
\label{fig:continuum}
\end{figure}

The last question we address is whether the single-thimble results becomes exact in
the continuum limit. Note that the continuum trajectory takes the value of $\hat g^2$
towards zero, so it is possible that in this limit the discrepancy vanishes as the 
weak coupling simulations suggests. As one approaches the continuum limit the location 
of the critical points and the contribution of their respective thimbles to observables, changes. To answer this question we performed a series of simulations with increasing 
values of $N$ while adjusting $\hat m$, $\hat \mu$, and $\hat g^2$ according to the
formulas in Section~\ref{sec:contlim}. We started with the strong coupling set of 
parameters at $N=4$ where the lattice spacing was taken to be $a=1$. The results for 
the particle density and the condensate are shown in \fig{fig:continuum}. The successive 
calculations with increasing values of $N=4, 8, 16, 32$, and $64$ clearly converge but
{\em not} to the exact result. Thus even in the continuum limit the subleading thimbles
have a non-vanishing contribution. 

\section{Conclusions}
\label{sec:conclusions}

In this paper we proposed an algorithm for the computation of single-thimble contribution 
to field theories. Our method has some advantages over previously proposed algorithms, in 
particular the fact that it relies only on integration of the flow in the numerically 
stable direction and it avoids flowing the points or the tangent vectors in the 
unstable direction of the flow (towards the critical point). The computationally costly 
part of the algorithm is the computation of the Jacobian of the upward flow which scales 
as $V^2$ in terms of memory footprint and $V^3$ in terms of floating-point operations, 
where $V$ is the spacetime volume. The computational cost can be reduced if the
Hessian is sparse or has a simple structure (as is the case for the model we used). 

We applied the algorithm to a simple, solvable one-site fermionic model and demonstrated 
its feasibility. The algorithm performs well despite some peculiarities of fermionic
model as the presence of singularities on the borders of the thimble (where the 
fermion determinant vanishes).

At weak coupling and high temperature there is a good agreement between the Monte 
Carlo calculation of the  one thimble contribution and the exact result. There, 
the semiclassical estimates for the contributions of other thimbles indicate they are small.

On the other hand, at strong coupling or low temperatures, the discrepancy between the 
one thimble and the exact results is noted numerically; semiclassical estimates suggest 
the contribution of other thimbles have the correct order of magnitude to fill in the gap.

A sizable contribution from other thimbles survive in the continuum limit.
Arguments have been put forward suggesting that for the continuum limit of field 
theories or systems with a thermodynamic limit simulations performed in one single 
thimble suffice. These arguments are based on the assumption that the theory defined 
on one thimble is in the same universality class as the theory defined over real 
variables (or, what is the same, over all thimbles). That way the contribution from 
other thimbles would be simply to renormalize the parameters of the one-thimble theory. 
Despite detecting a discrepancy between the one thimble and the exact result in 
physical observables our calculation does not bear on this issue since there is not 
concept of universality in $0+1$ dimensions. A test in higher dimensions, near the 
continuum limit, would help settle this question. Hopefully, the algorithm 
we describe in this paper will help  achieve this goal.

\acknowledgements

A.A. is supported in part by the National Science Foundation CAREER grant PHY-1151648. 
A.A. would like to thank Luigi Scorzato and Christian Schmidt for interesting 
discussions regarding the topic of this paper and gratefully acknowledges the hospitality 
of the Physics Department at the University of Maryland where part of this work was 
carried out. G.B and P.B  are supported by U.S. Department of Energy under Contract No. DE-FG02-93ER-40762.

\bibliography{thimblemc_final}

\end{document}